\documentclass[showpacs,amsmath,amssymb,twocolumn,nofootinbib]{revtex4}
\usepackage{graphicx}
\usepackage{epsfig}
\usepackage{float}
\usepackage{braket}
\usepackage{tikz}

\begin{document}

\allowdisplaybreaks[1]

\title{Question of quantum equivalence between Jordan frame and Einstein frame}

\author{Alexander Yu.~Kamenshchik}
\email{Alexander.Kamenshchik@bo.infn.it}
\affiliation{Dipartimento di Fisica e Astronomia and INFN, Via Irnerio 46, 40126 Bologna, Italy\\
and L.~D.~Landau Institute for Theoretical Physics of the
Russian Academy of Sciences,\\
Kosygin str.~2, 119334 Moscow, Russia}

\author{Christian F.~Steinwachs}
\email{christian.steinwachs@physik.uni-freiburg.de}
\affiliation{Physikalisches Institut, Albert-Ludwigs-Universit\"at Freiburg,\\
Hermann-Herder-Str.~3, 79104 Freiburg, Germany}


\begin{abstract}

In the framework of a general scalar-tensor theory, we investigate the equivalence between two different parametrizations of fields that are commonly used in cosmology  - the so-called Jordan frame and Einstein frame.
While it is clear that both parametrizations are mathematically equivalent at the level of the classical action, the question about their mathematical equivalence at the quantum level as well as their physical equivalence is still a matter of debate in cosmology. We analyze whether the mathematical equivalence still holds when the first quantum corrections are taken into account.
We explicitly calculate the one-loop divergences in both parametrizations by using the generalized Schwinger-DeWitt algorithm and compare both results. We find that the quantum corrections do not coincide off shell and hence induce an off shell dependence on the parametrization.
According to the equivalence theorem, the one-loop divergences should however coincide on shell. For a cosmological background, we show explicitly that the on shell equivalence is indeed realized by a nontrivial cancellation.  
\end{abstract}


\pacs{04.60.-m; 98.80.Qc; 11.10.Gh}
\maketitle


\section{Introduction}\label{sec1}

The cosmological models based on scalar fields nonminimally coupled to gravity \cite{Spok}--\cite{LindeNew} have recently become more popular again, in particular because of inflationary models, in which the inflation is driven by the nonminimally coupled standard model Higgs boson \cite{Shap}--\cite{ShapNew}.

The renormalization group running that connects the present low-energy vacuum of the standard model with the high-energy phase during inflation is based on quantum corrections and is essential for the numerical predictions of the nonminimal Higgs inflation model \cite{Shap1,Wil,BKKSS}.
While the one-loop running seems to be not accurate enough to ensure the agreement of these models with the latest LHC results \cite{AtlasResultHiggs,CMSResultHiggs} (a difference of roughly $10$ GeV between the predicted and measured values for the Higgs mass), it turned out that the two-loop running \cite{Shap2} brings the lower end of the predicted Higgs mass spectrum very close to the measured value of $M_{H}\sim126$ GeV \cite{AtlasResultHiggs, CMSResultHiggs}.

Although it is an appealing feature of these models that they are falsifiable and indeed produce numerical predictions that are in agreement with the latest results of the satellite PLANCK \cite{Planck} as well as with the recently announced Higgs mass \cite{AtlasResultHiggs, CMSResultHiggs}, one should first clarify the basic principles of the model before worrying about exact numerical values. Some of these principle questions have already been discussed recently, among them the multiplet nature of the standard model complex Higgs $SU(2)$ doublet \cite{Kaiser,LernerII,Hertzberg} and the question of unitarity for the high-energy phase during inflation \cite{Burgess1}--\cite{ Calmet:2013hia}. However, probably the most fundamental problem, connected to the question of the equivalence of different field parametrizations, still remains unsolved. The purpose of this paper is to address this problem.

In cosmological models with a scalar field nonminimally coupled to gravity two particular parametrizations of fields that are commonly used are denoted ``Jordan frame'' (JF) and ``Einstein frame'' (EF). 
There is an ongoing debate with quite a long history about the equivalence of these parametrizations, see e.g. \cite{Faraoni}--\cite{Domenech:2015qoa}. This debate can be subdivided into the question of mathematical and physical equivalence. We mainly focus on the mathematical aspects but comment in the conclusion also about possible physical implications.

While it is rather easy to check explicitly that the mathematical description in the two frames is equivalent at the tree level, it is not so obvious whether this equivalence will still hold at the quantum level. There has been much activity in analyzing the frame equivalence of cosmological observables and perturbations \cite{Catena:2006bd}--\cite{Prokopec:2014iya}. Although the cosmological perturbations are quantized, this is a ``mode-by-mode'' analysis and does not involve the quantum divergences that arise due to loop effects. Therefore, the analysis of cosmological perturbations does not answer the question whether the renormalization and the backreaction on the background, on which these perturbations propagate, depend on the chosen parametrization. 

This paper provides a natural application of the result obtained in our preceding paper \cite{we-paper1}, where we have calculated the one-loop divergences in the Jordan frame for a general scalar-tensor theory.
Here, we use this result in order to address the question of equivalence at the quantum level by explicitly calculating the first quantum corrections in the Einstein frame and Jordan frame parametrizations.

The strategy of our calculation can be summarized as follows: We choose to start in the Jordan frame parametrization. In the first calculation, we  compute the one-loop divergences directly in the Jordan frame. In the second calculation, we first transform the tree level action from the Jordan frame parametrization to the Einstein frame parametrization. Then we calculate the one-loop divergences in the Einstein frame parametrization and finally express the obtained quantum result again in the Jordan frame parametrization, in order to compare the two results obtained by quantizing in different frames.

The question of equivalence at the quantum level can be rephrased as the question whether the following diagram commutes or not.
\begin{figure}[h]
\begin{center}
\includegraphics[width=8cm,height=4cm]{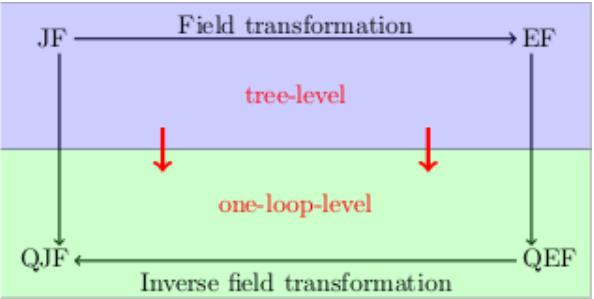}
\caption{\label{fig1}\small Transition between the (JF) and the (EF) at the tree level and Jordan frame (QJF) and Einstein frame (QEF) at the quantum level.}
\end{center}
\end{figure}

We find that the diagram does not commute off shell, which implies that already the first quantum corrections induce an off shell frame dependence. We suggest that this off shell frame dependence can be traced back to the ordinary definition of the effective action which is not covariant with respect to the configuration space of fields. 

The paper is structured as follows:

In Sec.~\ref{sec2}, we make use of the result \cite{we-paper1} for the one-loop divergences of a general scalar-tensor theory with a $O(N)$-symmetric scalar multiplet, calculated in the Jordan frame. By reducing this result to the single field case $N=1$, we obtain the result for the direct Jordan frame calculation.

In Sec.~\ref{sec3}, we perform the transformation of the Jordan frame tree level action to the Einstein frame and calculate the one-loop divergences in the Einstein frame by using the generalized Schwinger-DeWitt method \cite{DeWitt,Bar-Vil}.

In Sec.~\ref{sec4} we express the Einstein frame one-loop result obtained in Sec.~\ref{sec3} again in terms of the Jordan frame parametrization by applying the inverse transformation back to the Jordan frame. We then compare this with the one-loop divergences obtained by the direct Jordan frame calculations in Sec.~\ref{sec2}. We find that both quantum results do not coincide off shell which shows explicitly the frame dependence of the off shell one-loop divergences.

In Sec.~\ref{sec5} we investigate whether the obtained results are in agreement with the equivalence theorem \cite{Borchers}--\cite{Tyutin:2000ht}. According to this formal theorem, the one-loop divergences are guaranteed to coincide on shell. In the presence of gravity, the explicit realization of this on shell equivalence is nontrivial. Therefore we explicitly compare the Jordan frame and Einstein frame results on shell. Making use of the equations of motion, we show explicitly that the on shell equivalence is indeed realized in a nontrivial way for a simple cosmological background.

Finally, in Sect.~\ref{sec6} we conclude with a discussion of the obtained results and their possible implications for cosmology.

The explicit conformal transformation formulas for all expressions that appear in the calculations are collected in Appendix \ref{AppA}.
The rather long expressions for all the coefficients that enter the divergent one-loop contributions to the effective action in the Jordan frame and Einstein frame are presented in Appendixes \ref{AppB} and \ref{AppC} respectively. The technical details of the explicit one-loop calculation in the Einstein frame can be found in Appendix \ref{AppD}.


\section{Jordan Frame Calculation}\label{sec2}

We consider the action
\begin{align}
S = \int_{\cal M} \text{d}^4x \sqrt{g}\left(U\,R - \frac{1}{2}\,G\, g^{\mu\nu}\,\partial_{\mu}\Phi^{a}\partial_{\nu}\Phi_a-V\right)
\label{actionN}
\end{align}
for a general scalar-tensor theory of a nonminimally coupled $O(N)$-symmetric scalar multiplet $\phi^{a}$ with the modulus
\begin{align}
\varphi:=\sqrt{\delta_{ab}\Phi^a\Phi^b},\quad  a = 1,\cdots,N.
\label{varphi}
\end{align}
The field dependent couplings $U(\varphi)$, $G(\varphi)$ and $V(\varphi)$ are invariant with respect to the rotations in the isotopic $N$-dimensional space.\footnote{We choose to work with the conventions $\text{sign}(g_{\mu\nu})=+2$, $g:= det g_{\mu\nu}$, $R_{\mu\nu}:=R^{\alpha}_{\;\;\mu\alpha\nu}$, $R^{\alpha}_{\;\;\mu\beta\nu}:=\partial_{\beta}\,\Gamma^{\alpha}_{\;\;\mu\nu}-...$, $\Box:=g^{\mu\nu}\,\nabla_{\mu}\nabla_{\nu}$ }
In \cite{we-paper1} the divergent part of the one-loop effective action was calculated in a closed form by making use of the generalized Schwinger-DeWitt technique \cite{DeWitt,Bar-Vil}.
All the calculations were carried out consistently in the Jordan frame parametrization. We note that the calculations in the Jordan frame \cite{we-paper1} are especially important, because for a multiplet of scalar fields, such as e.g. the complex standard model Higgs doublet, the transition to the Einstein frame is not possible in general \cite{Kaiser}. Therefore, in such a case there is no corresponding Einstein frame.

Here, we reduce this general result to the case of a single scalar field $N=1$ with the action
\begin{align}
S^{\rm{J}} = \int_{\cal M} \text{d}^4x \sqrt{g}\left(U\,R - \frac{1}{2}\,G\, g^{\mu\nu}\,\partial_{\mu}\varphi\,\partial_{\nu}\varphi-V\right)\,,
\label{action}
\end{align}
and investigate whether the calculation of the one-loop divergences of (\ref{action}) parametrized in two different frames will lead to different results.

Using the {\tt MathTensor} package \cite{MathTensor} to reduce the general result obtained in \cite{we-paper1} to the single field case $N=1$, we obtain the one-loop divergences calculated in the Jordan frame
\begin{widetext}
\begin{align}
\Gamma_{1\rm{-loop}}^{\rm{div},\;\rm{J}}
={}&\frac{1}{32\pi^2(\omega-2)}\int\text{d}^4x\,\sqrt{g}\,\Bigg\{U_{1-\rm{loop}}^{\rm{J}}\,R+G_{1-\rm{loop}}^{\rm{J}}\,(\varphi_{,\,\nu}\varphi^{,\,\nu})+V_{1-\rm{loop}}^{\rm{J}}+\alpha_{1}^{\rm{J}}\,R_{\mu\nu}R^{\mu\nu}+\alpha_{2}^{\rm{J}}\,R^2+\alpha_{3}^{\rm{J}}\,R\,(\Box\,\varphi)\nonumber\\
&+\alpha_{4}^{\rm{J}}\,R\,(\varphi_{,\,\nu}\,\varphi^{,\,\nu})+\alpha_{5}^{\rm{J}}\,R^{\mu\nu}\,\varphi_{,\,\mu}\,\varphi_{,\,\nu}+\alpha_{6}^{\rm{J}}\,(\Box\,\varphi)^2+\alpha_{7}^{\rm{J}}\,(\Box\,\varphi)\,(\varphi_{,\,\nu}\,\varphi^{,\,\nu})+\alpha_{8}^{\rm{J}}\,(\varphi_{,\,\nu}\,\varphi^{,\,\nu})^2\Bigg\}\,.\label{JFEFFAction}
\end{align}
\end{widetext}
Here, $1/(\omega-2)$ is a pole in dimension.
The explicit form of the individual off shell coefficients $U_{1-\rm{loop}}^{\rm{J}}$, $G_{1-\rm{loop}}^{\rm{J}}$ and $\alpha_{i}^{\rm{J}},\quad i=1,..,8$ can be found in Appendix \ref{AppB}. In the main text, we only focus on the one-loop corrections to the off shell effective potential $V_{1-\rm{loop}}^{\rm{J}}$. This is not only the most important structure regarding cosmological applications: it is also the only structure that does not contain any space-time derivatives and thereby can never receive any contributions from other structures due to integration by parts. Since the basis in the space of independent invariants (scalar contractions of field operators, space-time derivatives and curvature terms) is -to some extent - a matter of choice, the off shell effective potential can serve as a unique indicator to test the off shell quantum equivalence of different frames in the following sense: It is already sufficient to show the nonequivalence of the off shell effective potential 
calculated in different frames, in order to show the frame dependence of the off shell quantum results. 
The off shell one-loop corrections to the potential in the Jordan frame are explicitly given by
\begin{widetext}
\begin{align}
\sqrt{g}\,V^{\rm{J}}_{1\text{-loop}}=&\sqrt{g}\,\Bigg\{\;V^2
   \left[2\, s^2\, \frac{\left(U'\right)^4}{U^4}-2\, s\,\frac{
   \left(U'\right)^2}{U^3}+\frac{5}{U^2}\right]+V V'
   \left[-8\, s^2\,\frac{
   \left(U'\right)^3}{U^3}+4\, s\,\frac{ U'}{U^2}\right]+2\,V\,V''\,s^2\,\frac{
   \left(U'\right)^2 }{U^2}\nonumber\\
&+\left(V'\right)^2 \left[8\, s^2\,\frac{
   \left(U'\right)^2}{U^2}-\frac{2 s}{U}\right]- 4 \,V' V''\,s^2\,\frac{ U'}{U}+\frac{1}{2} \left(V''\right)^2 s^2\Bigg\}\,.
   \label{V1loop}
\end{align}
\end{widetext}
Here, a prime denotes a derivative with respect to the Jordan frame field $\varphi$ and $s(\varphi)$ is a particular combination of the field dependent generalized potentials $G$ and $U$
\begin{align}
s=\frac{U}{G\,U+3\,(U')^2}\,.\label{suppfunct}
\end{align}
 A similar calculation in the Jordan frame for a single scalar field was already performed before in \cite{Shapiro}.\footnote{See also \cite{Alvarez:2014qca, Nojiri:2000ja, Bamba:2014mua} for one-loop calculations in the context of dilaton gravity.} As discussed in \cite{we-paper1}, a few discrepancies in several coefficients $\alpha_i$ remain when comparing the result obtained in  \cite{Shapiro} with the limiting case $N=1$ of the general result derived in \cite{we-paper1}. This, however, does not affect the most important structures and in particular not the off shell effective potential, so that the conclusion drawn here remains valid independent of these discrepancies.


\section{Einstein Frame Calculation}\label{sec3}

In the next step, we transform the tree-level action (\ref{action}) into the Einstein frame. In the absence of any additional matter fields apart from the scalar field, the Einstein frame is defined as the particular parametrization of fields $(\hat{g}_{\mu\nu},\,\hat{\varphi})$ in which (\ref{action}) formally resembles the ordinary Einstein-Hilbert action with a minimally coupled scalar field. The transformation procedure can be subdivided into two steps. First, we perform a conformal transformation of the metric field $g_{\mu\nu}\to\hat{g}_{\mu\nu}$ in order to remove the nonminimal coupling term. Afterwards, we perform an additional reparametrization of the scalar field $\varphi\to\hat{\varphi}$ such that the kinetic term acquires the standard canonically normalized form.

In order to find the explicit transformation law that connects the Jordan frame parametrization with the Einstein frame parametrization, we have to investigate how the nonminimal term in the tree-level action (\ref{action}) changes under a conformal transformation $g_{\mu\nu}=f\,\hat{g}_{\mu\nu}$ with a field dependent conformal factor $f(\varphi)$.
Using the general conformal transformation laws provided in Appendix \ref{AppA} we find
\begin{align}
 U\,\sqrt{g}\,R=\,U\sqrt{\hat{g}}\,\left(f\,\hat{R}+\frac{3}{2}\,f^{-1}\,f_{;\nu}\,f^{;\,\nu}-3\,f_{;\nu}^{\;\;\nu}\right)\,.
\label{uaction}
\end{align}
All quantities expressed in terms of the Einstein frame parametrization are denoted by a hat.
In order to remove the nonminimal coupling (keeping only the Einstein-Hilbert term which can be associated with the constant $U_{0}$) we have to choose
\begin{align}
 f=\frac{U_{0}}{U}\,.
\label{conformalfac}
\end{align}
Under this conformal transformation, the potential and the kinetic term in (\ref{action}) will be simply rescaled by powers of the conformal factor (\ref{conformalfac}) due to the transformation of the volume element,
\begin{align}
\sqrt{g}\,V=&\sqrt{\hat{g}}\,\left(\frac{U_{0}}{U}\right)^2\,V=\sqrt{\hat{g}}\,\hat{V}\,,\label{redefPot}\\
\sqrt{g}\,G\,g^{\mu\nu}\,\varphi_{,\,\mu}\,\varphi_{,\,\nu}=&\sqrt{\hat{g}}\,\left(\frac{U_{0}}{U}\right)\,G\,\hat{g}^{\mu\nu}\,\varphi_{,\,\mu}\,\varphi_{,\,\nu}\label{redfkin}\,.
\end{align}
In the last equality of (\ref{redefPot}) we have absorbed the conformal factor by a redefinition of the potential
\begin{align}
\hat{V}(\varphi):= U_{0}^2\;\frac{V(\varphi)}{U^2(\varphi)}\;.
\label{EFPot}
\end{align}
Integration by parts of the last term in (\ref{uaction}) with the choice (\ref{conformalfac}) leads to
\begin{align}
U\,\sqrt{g}\,R=&U_{0}\,\sqrt{\hat{g}}\left[\hat{R}-\frac{3}{2}\,\left(\frac{U'}{U}\right)^2\,\varphi_{;\,\mu}\,\varphi^{\;\,\mu}\right]\,.
\end{align}
Thus, the right-hand side of (\ref{redfkin}) receives an extra contribution, leading to the transformed kinetic term
\begin{align}
-\frac{1}{2}\,\sqrt{g}\,G\,g^{\mu\nu}\,\varphi_{,\,\mu}\,\varphi_{,\,\nu}=-\frac{1}{2}\,M\,\sqrt{\hat{g}}\,\hat{g}^{\mu\nu}\,\varphi_{,\,\mu}\,\varphi_{,\,\nu}\,,
\label{mkin}
\end{align}
with the field dependent quantity
\begin{align}
M(\varphi):=\left(\frac{U_{0}}{U}\right)\left(\,\frac{G\,U+3\,(U')^2}{U}\right)\,.
\end{align}
In order to normalize the kinetic term, we can again make use of the fact that $\varphi$ is just a configuration space variable and perform an additional field reparametrization
\begin{align}
 \varphi\rightarrow\hat{\varphi}\,.\label{FieldTransformation}
\end{align}
The condition for the field transformation law is fixed by (\ref{mkin}).
We must therefore find a solution $\hat{\varphi}$ to the equation
\begin{align}
 M(\varphi)\,\hat{g}^{\mu\nu}\varphi_{,\,\mu}\varphi_{,\,\nu}= \hat{g}^{\mu\nu}\,\hat{\varphi}_{,\,\mu}\hat{\varphi}_{,\,\nu}\,.
\end{align}
This can equivalently be written as the condition
\begin{align}
 \left(\frac{\partial\,\hat{\varphi}}{\partial \varphi}\right)^2=\left(\frac{U_{0}}{U}\right)\left(\,\frac{G\,U+3\,(U')^2}{U}\right)\,.
\label{nonlinfieldtrans}
\end{align}
Later, we will also need the inverse relation
\begin{align}
\left(\frac{\partial \varphi}{\partial\,\hat{\varphi}}\right)^2=\left(\frac{U}{U_{0}}\right)\left(\,\frac{U}{G\,U+3\,(U')^2}\right)\,.\label{InverseFieldTrafo}
\end{align}
After the conformal transformation with (\ref{conformalfac}), the field transformation (\ref{InverseFieldTrafo}) and the redefinition of the potential (\ref{EFPot}) can be used to finally express the action (\ref{action}) in terms of the Einstein frame parametrization
\begin{align}
\hat{S} = \int_{\cal M} \text{d}^4x \sqrt{\hat{g}}\left(U_{0}\,\hat{R} - \frac{1}{2}\, \hat{g}^{\mu\nu}\,\partial_{\mu}\hat{\varphi}\,\partial_{\nu}\hat{\varphi}-\hat{V}\right)\,.
\label{actionEF}
\end{align}
Using again the generalized Schwinger-DeWitt algorithm \cite{Bar-Vil}, we calculate the one-loop divergences for the action (\ref{actionEF}) in the Einstein frame parametrization.
We perform the calculation independently in two different ways. First, we repeat the explicit calculation that was done already in \cite{we-renorm}.
A brief summary of the details of that calculation can be found in Appendix \ref{AppD}.
Second, we use the general Jordan frame result for the $O(N)$-symmetric multiplet of scalar fields, obtained in \cite{we-paper1}, reduce it to the single field case $N=1$ and set in addition $U=U_{0}$ and $G=1$. Since the general Jordan frame result contains the Einstein frame result as a special case, both calculations should lead to the same result. Indeed, we independently find for both calculations the same result
\begin{widetext}
\begin{align}
\hat{\Gamma}_{1\rm{-loop}}^{\rm{div},\;\rm{E}}
={}&\frac{1}{32\pi^2(\omega-2)}\int\text{d}^4x\,\sqrt{\hat{g}}\,\Bigg\{\frac{43}{60}\,\hat{R}_{\mu\nu}\hat{R}^{\mu\nu}+\frac{1}{40}\,\hat{R}^2-\frac{1}{6}\,\hat{R}\,(\partial^2_{\hat{\varphi}}\hat{V})+\frac{1}{2}\,(\partial^2_{\hat{\varphi}}\hat{V})^2-U_{0}^{-1}\,\left[\frac{1}{3}\,\hat{R}\,(\hat{\varphi}_{,\,\nu}\hat{\varphi}^{,\,\nu})\right.\nonumber\\
&\left.+\frac{13}{3}\,\hat{R}\,\hat{V}+2\,(\partial^2_{\hat{\varphi}}\hat{V})\,(\hat{\varphi}_{,\,\nu}\hat{\varphi}^{,\,\nu})+2\,(\partial_{\hat{\varphi}}\hat{V})^2\right]+U_{0}^{-2}\,\left[5\,\hat{V}^2+\frac{5}{4}\,(\hat{\varphi}_{,\,\nu}\hat{\varphi}^{,\,\nu})^2+\,\hat{V}\,(\hat{\varphi}_{,\,\nu}\hat{\varphi}^{,\,\nu})\right]\Bigg\}\,.\label{EffActionEF}
\end{align}
\end{widetext}
Here, the derivative $\partial_{\hat{\varphi}}$ has to be computed with respect to the Einstein frame field $\hat{\varphi}$.
This result coincides with the one already obtained in \cite{we-renorm}.\footnote{Except for the $\hat{V}^2$ structure: instead of the correct $5\,\hat{V}^2$ in (\ref{EffActionEF}), in \cite{we-renorm}, there was a wrong prefactor $5/2\,\hat{V}^2$, which was already noted by the authors of \cite{we-renorm} in \cite{we2}.} In particular, we find for the divergent one-loop contribution to the effective potential calculated in the Einstein frame parametrization 
\begin{align}
 \sqrt{\hat{g}}\,\hat{V}^{\rm{E}}_{1-\rm{loop}}=\sqrt{\hat{g}}\,\left(\frac{1}{2}\,(\partial^2_{\hat{\varphi}}\hat{V})^2-\frac{2}{U_0}\,(\partial_{\hat{\varphi}}\hat{V})^2+\frac{5}{U_{0}^2}\,\hat{V}^2\right)\,.
\end{align}


\section{Einstein Frame Quantum Result Expressed in the Jordan Frame}\label{sec4}

Finally, we express the one-loop off shell divergences (\ref{EffActionEF}), derived from the tree level action in the Einstein frame parametrization (\ref{actionEF}), in terms of the original Jordan frame parametrization.
Thus, we need to perform the inverse transformation
\begin{align}
\hat{\Gamma}_{1\rm{-loop}}^{\rm{div},\;\rm{E}}\to\Gamma_{1\rm{-loop}}^{\rm{div},\;\rm{E}}\,,
\end{align}
in order to the explicit expression for $\Gamma_{1\rm{-loop}}^{\rm{div},\;\rm{E}}$. Here, quantities without a hat (with a hat) are represented in the Jordan frame (Einstein frame) parametrization and the superscript $J$ ($E$) denotes the Jordan frame (Einstein frame) parametrization that was used to calculate the one-loop divergences.
As we have already mentioned in Sec.~\ref{sec2}, it is sufficient to show that the two effective off shell potentials calculated in the Jordan frame and in the Einstein frame do not coincide when both are expressed in terms of the Jordan frame parametrization. The explicit form of the remaining coefficients $U_{1-\rm{loop}}^{\rm{E}}$, $G_{1-\rm{loop}}^{\rm{E}}$ and $\alpha_{i}^{\rm{E}},\quad i=1,...,8$ can be found in Appendix \ref{AppC}. Since all other invariant scalar structures, except for the potential, involve space-time derivatives, the inverse conformal and scalar field transformation of these structures will again involve space-time derivatives and therefore cannot lead to any contributions to the off shell effective potential. Thus, for the off shell comparison, it is sufficient to consider the inverse transformation of the divergent one-loop contributions to the effective potential: We simply have to express $\sqrt{\hat{g}}\,\hat{V}^{\rm{E}}_{1-\rm{loop}}$ in terms of the Jordan frame 
parametrization of 
fields $(g_{\mu\nu},\,\varphi)$.

The relevant transformation rules involve the square root of the metric determinant as well as the potential and its derivatives.
Making use of (\ref{cofdet}), (\ref{EFPot}) and (\ref{InverseFieldTrafo}), we find
\begin{align}
\sqrt{\hat{g}}={}&\frac{U^2}{U_{0}^2}\,\sqrt{g}\,,\\
\hat{V}=&\frac{U_{0}^2}{U^2}\,V\,,\\
\frac{\partial\hat{V}}{\partial\,\hat{\varphi}}={}&
\frac{U_{0}^{3/2}}{U^2}\,\frac{(V'\,U-2\,U'\,V)}{\sqrt{G\,U+3\,U'^2}}\;,\label{FirstDervPot}
\end{align}
\begin{widetext}
\begin{align}
\frac{\partial^2\,\hat{V}}{\partial\hat{\varphi}^2}={}&\frac{U_{0}}{U^2\,\left[G\,U+3\,(U)'^2\right]^2}\,\Bigg[12\,V\,(U')^4-9\,U\,(U')^3\,V'-3\,U^2\,U'\,V'\,U''+3\,U^2\,(U')^2\,V''\nonumber\\
&+G\,\Big(5\,U\,V\,(U')^2-\frac{7}{2}\,U^2\,U'\,V'-2\,U^2\,V\,U''+U^3\,V''\Big)+G'\,\left(U^2\,U'\,V-\frac{1}{2}\,U^3\,V'\right)\Bigg]\,.\label{SecondDervPot}
\end{align}

The divergent one-loop contribution to the off shell effective potential, calculated in the Einstein frame parametrization and expressed in terms of the Jordan frame parametrization, then reads 
\begin{align}
\sqrt{g}\,V_{1-\rm{loop}}^{\rm{E}}={}&\sqrt{g}\Bigg\{V^2 \left[
\,s^4 \left(
\frac{6 G' \left(U'\right)^3
   U''}{U^3}-\frac{3 G'
   \left(U'\right)^5}{U^4}+\frac{\left(G'\right)^2
   \left(U'\right)^2}{2 U^2}-\frac{18 \left(U'\right)^6
   U''}{U^5}+\frac{18 \left(U'\right)^4
   \left(U''\right)^2}{U^4}+\frac{9 \left(U'\right)^8}{2
   U^6}
   \right)\right.\nonumber\\
   &\quad{}+s^3 \left(
   -\frac{2 G' U'
   U''}{U^2}+\frac{5 G'
   \left(U'\right)^3}{U^3}+\frac{36 \left(U'\right)^4
   U''}{U^4}-\frac{12 \left(U'\right)^2
   \left(U''\right)^2}{U^3}-\frac{15
   \left(U'\right)^6}{U^5}
   \right)\nonumber\\
   &\quad{}+s^2 \left.\left(
   -\frac{10
   \left(U'\right)^2 U''}{U^3}+\frac{25
   \left(U'\right)^4}{2 U^4}+\frac{2
   \left(U''\right)^2}{U^2}
   \right)
   -s\,\frac{8\left(U'\right)^2}{U^3}
   +\frac{5}{U^2}
   \right]\nonumber\\
   &+V V'
   \left[s^4 \left(
   \frac{-6 G' \left(U'\right)^2
   U''}{U^2}+\frac{3 G' \left(U'\right)^4}{U^3}-\frac{
   \left(G'\right)^2 U'}{2\,U}+\frac{18 \left(U'\right)^5
   U''}{U^4}-\frac{18 \left(U'\right)^3 \left(U''\right)^2}{U^2}-\frac{9
   \left(U'\right)^7}{2\,U^5}
   \right)\right.\nonumber\\
   &\left.\quad{}+s^3\left(
   \frac{G' U''}{U}-\frac{6 G' \left(U'\right)^2}{U^2}-\frac{39
   \left(U'\right)^3 U''}{U^3}+\frac{6 U'
   \left(U''\right)^2}{U^2}+\frac{18 \left(U'\right)^5}{U^4}\right)+s^2 \left(\frac{7 U' U''}{U^2}-\frac{35
   \left(U'\right)^3}{2\,U^3}\right)+ s\,\frac{8
   U'}{U^2}
   \right]\nonumber\\
   &+V V'' \left[
   s^3
   \left(\frac{G' U'}{U}+6\frac{\left(U'\right)^2
   U''}{U^2}-\frac{3 \left(U'\right)^4}{U^3}\right)+s^2 \left(\frac{5
   \left(U'\right)^2}{U^2}-\frac{2
   U''}{U}\right)\right]\nonumber\\
   &+\left(V'\right)^2 \left[s^4
   \left(\frac{3 G' U' U''}{2 U}-\frac{3 G'
   \left(U'\right)^3}{4 U^2}+\frac{\left(G'\right)^2}{8}-\frac{9
   \left(U'\right)^4 U''}{2 U^3}+\frac{9 \left(U'\right)^2
   \left(U''\right)^2}{2 U^2}+\frac{9 \left(U'\right)^6}{8
   U^4}\right)\right.\nonumber\\
   &\quad{}+s^3\left. \left(\frac{7 G' U'}{4
   U}+\frac{21 \left(U'\right)^2 U''}{2 U^2}-\frac{21
   \left(U'\right)^4}{4 U^3}\right)+\frac{49 s^2
   \left(U'\right)^2}{8 U^2}-s\,\frac{2 }{U}\right]\nonumber\\
   &+V'
   V'' \left[s^3 \left(-\frac{G'}{2}-\frac{3 U'
   U''}{U}+\frac{3 \left(U'\right)^3}{2 U^2}\right)-s^2\,\frac{7
    U'}{2 U}\right]+s^2\,\frac{1}{2}\,\left(V''\right)^2 \Bigg\}\;.\label{VAOneLoop}
\end{align}
\end{widetext}
This quantity should now be compared with the divergent one-loop contribution to the effective potential directly calculated in the Jordan frame $V_{1-\rm{loop}}^{\rm{J}}$ that was given in (\ref{V1loop}).
Defining the difference $\Delta V_{1\rm{-loop}}:=V^{\rm{E}}_{1-\rm{loop}}-V^{\rm{J}}_{1-\rm{loop}}$, we find $\Delta  V_{1\rm{-loop}}\neq0$.
Thus, we have shown that the off shell one-loop divergences calculated in both parametrizations do not lead to the same result. Therefore, already the first quantum corrections induce an off shell frame dependence.


\section{On-shell Comparison of Jordan vs. Einstein Frame Quantization}\label{sec5}

The explicit on shell comparison of $\Gamma^{J}_{1-\rm{loop}}$ and $\Gamma^{E}_{1-\rm{loop}}$ found in Secs.~\ref{sec2} and \ref{sec4} is difficult.
The equations of motion in the Jordan frame are easily obtained by the first variation of the Jordan frame action (\ref{action}). Variation with respect to $g_{\mu\nu}$ gives the field equation for the metric field
\begin{widetext}
 \begin{align}
 R_{\alpha\beta}-\frac{1}{2}\,g_{\alpha\beta}\,R=\left(\frac{G+2\,U''}{2\,U}\right)\,\varphi_{,\alpha}\,\varphi_{,\beta}-\left(\frac{G+4\,U''}{4\,U}\right)\,g_{\alpha\beta}\,(\nabla\varphi)^2+\frac{U'}{U}\,\varphi_{;\,\alpha\beta}-\frac{U'}{U}\,g_{\alpha\beta}\,\Box\varphi-\frac{1}{2}\,g_{\alpha\beta}\,\frac{V}{U}\,.\label{FieldEqMetric}
 \end{align}
 \end{widetext}
 Contraction of (\ref{FieldEqMetric}) with $g^{\alpha\beta}$, $R^{\alpha\beta}$, $\varphi^{,\,\alpha}\,\varphi^{,\,\beta}$ and $\varphi^{;\,\alpha\beta}$ yields four scalar equations that can be used to reduce the number of independent structures in the effective actions
 \begin{widetext}
 \begin{align}
 R={}&\left(\frac{G+6\,U''}{2\,U}\right)\,(\nabla\varphi)^2+3\,\frac{U'}{U}\,\Box\varphi+2\,\frac{V}{U},\label{ScalarContrFieldEq1}\\
 R^{\mu\nu}R_{\mu\nu}={}&\frac{R^2}{2}+\left(\frac{G+2\,U''}{2\,U}\right)R^{\mu\nu}\,\varphi_{,\,\mu}\,\varphi_{,\,\nu}-\left(\frac{G+4\,U''}{4\,U}\right)\,R\,(\nabla\varphi)^2+\frac{U'}{U}\,R^{\mu\nu}\varphi_{;\,\mu\nu}-\frac{U'}{U}\,R\,\Box\varphi-\frac{V}{2\,U}R,\label{ScalarContrFieldEq2}\\
 R^{\mu\nu}\varphi_{,\,\mu}\varphi_{,\,\nu}={}&\frac{R}{2}\,(\nabla\varphi)^2+\frac{G}{4\,U}(\nabla\varphi)^4+\frac{U'}{U}\,\varphi^{;\,\mu\nu}\,\varphi_{,\,\mu}\varphi_{,\,\nu}-\frac{U'}{U}\,\Box\varphi\,(\nabla\varphi)^2-\frac{1}{2}\,\frac{V}{U}\,(\nabla\varphi)^2,\\
 R^{\mu\nu}\varphi_{;\,\mu\nu}={}&\frac{R}{2}\,\Box\,\varphi+\left(\frac{G+2\,U''}{2\,U}\right)\varphi^{;\,\mu\nu}\varphi_{,\mu}\varphi_{,\nu}-\left(\frac{G+4\,U''}{4\,U}\right)\Box\,\varphi(\nabla\varphi)^2+\frac{U'}{U}\left[\varphi^{;\,\mu\nu}\varphi_{;\,\mu\nu}-(\Box\varphi)^2\right]-\frac{V}{2\,U}\,\Box\,\varphi\,.
 \end{align}
 \end{widetext}
  Variation with respect to $\varphi$ yields one additional scalar equation, the Klein-Gordon equation for the scalar field
 \begin{align}
 \Box \varphi=-\frac{U'}{G}\,R-\frac{1}{2}\,\frac{G'}{G}\,(\nabla\varphi)^2+\frac{V'}{G}\,.\label{FieldEqScalar}
 \end{align}
 These formulas show that the equations of motion relate scalar invariants containing space-time derivatives with scalar invariants containing no space-time derivatives. Therefore, the coefficients of the invariants containing space-time derivatives will contribute in a nontrivial way to the one-loop divergences of the on shell effective potential. Ultimately, these on shell contributions will alter the comparison between the on shell effective potential calculated in the Jordan and Einstein frame and should lead to an on shell frame independence as suggested by the equivalence theorem \cite{Borchers}--\cite{Tyutin:2000ht}.
 
 The process of explicitly eliminating the dependent structures by the equations of motion in the particular model under consideration is rather tedious. The system is nonlinear in the invariants and requires us to alternately use (\ref{ScalarContrFieldEq1})--(\ref{FieldEqScalar}) and integration by parts in an iterative procedure several times, in order to reduce the different invariant structures to a minimal independent set. 
 
 We will, however, provide a detailed calculation and prove the on shell frame independence explicitly for the special case of a canonically normalized $G(\varphi)=1$, constant background scalar field $\nabla\varphi=0$. This particular background implies that no structure that involves a derivative of the scalar field can appear in the divergent one-loop contribution to the effective action (\ref{JFEFFAction}). Moreover, from the form of the  transformation to the Einstein frame scalar field (\ref{InverseFieldTrafo}), it is clear that this also implies $\hat{\nabla}\hat{\varphi}=0$. Thus, the one-loop divergences calculated in the Jordan frame (\ref{JFEFFAction}) and in the Einstein frame (\ref{EffActionEF}) reduce to the following structures respectively:
 \begin{widetext}
 \begin{align}
 \Gamma^{\rm{div,\,J}}_{1\rm{-loop}}[g,\varphi]=&\frac{1}{32\,\pi^2\,(\omega-2)}\,\int\text{d}^4x\,\sqrt{g}\left\{U^{\rm{J}}_{1\rm{-loop}}\,R+V^{\rm{J}}_{1\rm{-loop}}+\alpha_{1}^{\rm{J}}\,R_{\mu\nu}R^{\mu\nu}+\alpha_{2}^{\rm{J}}\,R^2\right\}\,,\label{EAJFConstPhi}\\
 \hat{\Gamma}^{\rm{div,\,E}}_{1\rm{-loop}}[\hat{g},\hat{\varphi}]=&\frac{1}{32\,\pi^2\,(\omega-2)}\,\int\text{d}^4x\,\sqrt{\hat{g}}\left\{\hat{U}^{\rm{E}}_{1\rm{-loop}}\,\hat{R}+\hat{V}^{\rm{E}}_{1\rm{-loop}}+\hat{\alpha}_{1}^{\rm{E}}\,\hat{R}_{\mu\nu}\hat{R}^{\mu\nu}+\hat{\alpha}_{2}^{\rm{E}}\,\hat{R}^2\right\}\,.\label{EAEFConstPhi}
\end{align}
\end{widetext}
From the Jordan frame equations of motion (\ref{ScalarContrFieldEq1})--(\ref{ScalarContrFieldEq2}) and (\ref{FieldEqScalar}), we obtain in the $\nabla\varphi=0$ case the following identities.
The trace of the equations of motion for the metric field $g_{\mu\nu}$ reduces to
\begin{align}
 R\overset{\wedge}{=}2\,\frac{V}{U},\,
\end{align}
while the Klein-Gordon equation for the scalar field yields
\begin{align}
 R\overset{\wedge}{=}\frac{V'}{U'}\,.
\end{align}
Combining these two equations leads to a relation that allows us to express $V$ in terms of $U$,
\begin{align}
V\overset{\wedge}{=}U^2\label{OnShellPot}\,.
\end{align}
Inserting this again in the trace equation, we find
\begin{align}
 R\overset{\wedge}{=}2\,U\,\label{OnShellRicciScalar}.
\end{align}
Finally, using these results in (\ref{ScalarContrFieldEq2}), we find
\begin{align}
R_{\mu\nu}R^{\mu\nu}\overset{\wedge}{=}U^2\label{OnShellRicciTensorSquared}\,,
\end{align}
where the wedge over the equality sign indicates that we have made use of the equations of motion.
The conformal transformations (\ref{ConfRicciScalar}) and (\ref{ConfRicciTensorSquare}) become simple scaling relations since the additional derivative structures are absent in the case of the constant background scalar field,
 \begin{align}
  \hat{R}=&\frac{U_0}{U}\,R\,+\rm{terms}[\nabla\varphi],\\
  \hat{R}_{\mu\nu}\hat{R}^{\mu\nu}=&\left(\frac{U_0}{U}\right)^2\,R_{\mu\nu}R^{\mu\nu}+\rm{terms}[\nabla\varphi]\,.
 \end{align}
It remains to express the Einstein frame coefficients $\hat{U}^{\rm{E}}_{1\rm{-loop}}$, $\hat{V}^{\rm{E}}_{1\rm{-loop}}$ $\hat{\alpha}_{1}^{\rm{E}}$, $\hat{\alpha}_{2}^{\rm{E}}$ in terms of the Jordan frame field $\varphi$, in order to find $U^{\rm{E}}_{1\rm{-loop}}$, $V^{\rm{E}}_{1\rm{-loop}}$, $\alpha_{1}^{\rm{E}}$, $\alpha_{2}^{\rm{E}}$. For $G(\varphi)=1$, the suppression function (\ref{suppfunct}) becomes
\begin{align}
  s(\varphi)=\frac{U}{U+3\,(U')^2}\,.
\end{align}
The Einstein frame coefficients $\hat{U}^{\rm{E}}_{1\rm{-loop}}$, $\hat{V}^{\rm{E}}_{1\rm{-loop}}$ $\hat{\alpha}_{1}^{\rm{E}}$, $\hat{\alpha}_{2}^{\rm{E}}$ only involve the Einstein frame potential $\hat{V}$ and first and second derivatives thereof $\partial_{\hat{\varphi}}\hat{V}$ and $\partial^2_{\hat{\varphi}}\hat{V}$,
\begin{align}
 \hat{U}^{\rm{E}}_{1\rm{-loop}}&=-\frac{1}{6}\,(\partial^2_{\hat{\varphi}}\hat{V})-\frac{13}{3}\frac{\hat{V}}{U_0}\,,\\
 \hat{V}^{\rm{E}}_{1\rm{-loop}}&=\frac{1}{2}\,(\partial^2_{\hat{\varphi}}\hat{V})^2-\frac{2}{U_0}\,(\partial_{\hat{\varphi}}\hat{V})^2+\frac{5}{U_{0}^2}\,\hat{V}^2\,,\\
 \hat{\alpha}_{1}^{\rm{E}}&=\frac{43}{60}\,,\\
 \hat{\alpha}_{2}^{\rm{E}}&=\frac{1}{40}\,.
\end{align}
 With the definition (\ref{EFPot}) and the on shell relation (\ref{OnShellPot}), we find
\begin{align}
 \hat{V}=\frac{U_0^2}{U^2}\,V\overset{\wedge}{=}U_0^2\,.
\end{align}
Therefore, the relevant transformation formulas (\ref{FirstDervPot})--(\ref{SecondDervPot}) for the derivatives $\partial_{\hat{\varphi}}\hat{V}$ and $\partial^2_{\hat{\varphi}}\hat{V}$ imply $\partial_{\hat{\varphi}}\hat{V}\overset{\wedge}{=}0$ and $\partial^2_{\hat{\varphi}}\hat{V}\overset{\wedge}{=}0$.
The on shell Einstein frame coefficients expressed in the Jordan frame parametrization become
\begin{align}
 U^{\rm{E}}_{1\rm{-loop}}&\overset{\wedge}{=}-\frac{13}{3}\,U_0\,\label{OnShellU1LoopEF},\\
 V^{\rm{E}}_{1\rm{-loop}}&\overset{\wedge}{=}5\,U_0^2\,,\\
 \alpha_{1}^{\rm{E}}&\overset{\wedge}{=}\frac{43}{60}\,,\\
 \alpha_{2}^{\rm{E}}&\overset{\wedge}{=}\frac{1}{40}\label{OnShellAlpha2EF}\,.
\end{align}
Combining (\ref{OnShellU1LoopEF})--(\ref{OnShellAlpha2EF}) and (\ref{OnShellPot})--(\ref{OnShellRicciTensorSquared}), we obtain the Einstein frame on shell one-loop divergences (\ref{EAEFConstPhi}) expressed in terms of the Jordan frame parametrization
 \begin{align}
 \Gamma^{\rm{div,\,E}}_{1\rm{-loop}}[g,\varphi]\overset{\wedge}{=}\frac{1}{32\,\pi^2\,(\omega-2)}\,\int\text{d}^4x\,\sqrt{g}\,\left(-\frac{57}{20}\,U^2\right)\label{EFEffActionOnShell}\,.
\end{align}
In order to compare this with the on shell version of the one-loop divergences calculated in the Jordan frame (\ref{EAJFConstPhi}), we have to find the on shell values of the coefficients (\ref{V1loop}), (\ref{U1loop}) and (\ref{Alpha1JF1L})--(\ref{Alpha2JF1L}). Using again the relations (\ref{OnShellPot})--(\ref{OnShellRicciTensorSquared}) and $G(\varphi)=1$, we find
\begin{widetext}
 \begin{align}
 U^{\rm{J}}_{1\rm{-loop}}&\overset{\wedge}{=}-\frac{U \left\{U^2 \left[6 \left(U''\right)^2+U''+13\right]+117
   \left(U'\right)^4+7 U \left(U'\right)^2 \left(10-3
   U''\right)\right\}}{3 \left(3 \left(U'\right)^2+U\right)^2}\label{U1LoopJFOnShell}\,,\\
 V^{\rm{J}}_{1\rm{-loop}}&\overset{\wedge}{=}\frac{U^2 \left\{U^2 \left[2 \left(U''\right)^2+5\right]+47
   \left(U'\right)^4+4 U \left(U'\right)^2 \left(7-2
   U''\right)\right\}}{\left(3 \left(U'\right)^2+U\right)^2}\,,\\
 \alpha_{1}^{\rm{J}}&\overset{\wedge}{=}\frac{2 \left(U'\right)^2}{3 \left(U'\right)^2+U}+\frac{43}{60}\,,\\
 \alpha_{2}^{\rm{J}}&\overset{\wedge}{=}\frac{U^2 \left[60 \left(U''\right)^2+20 U''+3\right]-213
   \left(U'\right)^4-2 U \left(U'\right)^2 \left[90
   U''+71\right]}{120 \left(3 \left(U'\right)^2+U\right)^2}\label{Alpha21LoopJFOnShell}\,.
\end{align}
\end{widetext}
Combining the on shell coefficients (\ref{U1LoopJFOnShell})--(\ref{Alpha21LoopJFOnShell}) with the corresponding on shell structures (\ref{OnShellPot})--(\ref{OnShellRicciTensorSquared}), we obtain the on shell Jordan frame one-loop divergences (\ref{EAJFConstPhi})
\begin{align}
 \Gamma^{\rm{div,\,J}}_{1\rm{-loop}}[g,\varphi]\overset{\wedge}{=}&\frac{1}{32\,\pi^2\,(\omega-2)}\,\int\text{d}^4x\,\sqrt{g}\Bigg\{
  -\frac{57}{20} U^2\Bigg\}\,.\label{JFEffActionOnShell}
\end{align}
Comparing this result with (\ref{EFEffActionOnShell}), we indeed find that the divergent parts of the two effective actions do coincide on shell.
In particular this reveals how a very nontrivial cancellation between the complicated coefficients (\ref{U1LoopJFOnShell})--(\ref{Alpha21LoopJFOnShell}) ensures the on shell coincidence of the one-loop divergences (\ref{EFEffActionOnShell}) and (\ref{JFEffActionOnShell}) explicitly. This is in perfect agreement with the formal statement of the equivalence theorem \cite{Borchers}--\cite{Tyutin:2000ht}.\footnote{The on shell coincidence of the one-loop effective action, calculated in the Jordan frame and in the Einstein frame, was also investigated in \cite{Bamba:2014mua}.} Note that the classical background, representing the solution to the equations of motion with $V=U^2$ and $\nabla\varphi=0$ corresponds to the maximally symmetric de Sitter space-time.


\section{Conclusion}\label{sec6}

We have calculated the divergent part of the one-loop effective action for a general scalar-tensor theory involving a single scalar field nonminimally coupled to gravity in the Jordan frame and Einstein frame parametrizations. The corresponding results were obtained by applying the Schwinger-DeWitt method \cite{DeWitt,Bar-Vil} separately to the same classical action expressed in these two different parametrizations of fields. After the calculation of the one-loop divergences in both parametrizations, we have expressed both results in the Jordan frame parametrization in order to compare them. We have found that the two off shell results do not coincide. This implies that the classical equivalence between the two frames is destroyed already by the first quantum corrections and shows that the off shell one-loop divergences indeed depend on the parametrization.
  
By formal arguments, the equivalence theorem states that the quantum corrections in both frames should coincide on shell. In the presence of gravity, including graviton loops, the explicit realization of this on shell equivalence is however nontrivial. Therefore, we have carried out an explicit on shell analysis for the case of a constant background scalar field. In agreement with the equivalence theorem \cite{Borchers}--\cite{Tyutin:2000ht}, we have found that the on shell equivalence is indeed established by a nontrivial cancellation of contributions from different structures. A similar coincidence of on shell one-loop results for another model was found in \cite{Bamba:2014mua}. 

Finally, we discuss some points that go beyond the scope of this paper.

In this paper, we have focused on the prelogarithmic coefficients of the one-loop divergences. There is the additional problem of the conformal anomaly that also affects the logarithmic structure \cite{BKS}. This could be a hint that the off shell parametrization dependence is not only connected with the frame dependent calculation of the counterterms but in addition with the procedure of renormalization itself: see e.g. \cite{BMSS} for a discussion of this point.

As we have mentioned already in the introduction, the equivalence between the Jordan and Einstein frame is a controversial topic in cosmology. For the particular model of nonminimal Higgs inflation \cite{Shap}--\cite{ShapNew}, it was recently claimed in \cite{George:2013iia} that no indication for a frame dependence of the off shell one-loop effective potential has been found. The result of \cite{George:2013iia} therefore seems to be in contradiction with the off shell quantum parametrization dependence found in this paper. In \cite{George:2013iia}, the one-loop Coleman-Weinberg potential was calculated, including neither the contributions of Higgs loops nor of graviton loops. Such an approach might seem to be justified in the effective field theoretical framework of nonminimal Higgs inflation with a strong nonminimal coupling $\xi$. In this case, the Higgs propagator is suppressed by powers of the function (\ref{suppfunct}) that scales like $s\sim1/6\xi$ for high energies and graviton loops are suppressed by 
powers of the Planck mass. 
Such an analysis seems, however, to essentially miss the complicated and interesting part of the calculation: the graviton-scalar mixing that is a consequence of the nonminimal coupling to gravity. Instead, we have presented a fully (space-time) covariant calculation of the one-loop divergences that contribute to the effective potential, including graviton loops. We performed the calculation for a general background $(\bar{g}_{\mu\nu},\,\bar{\varphi})$, a general potential $V(\varphi)$ and a general nonminimal coupling $U(\varphi)$. We believe that the inclusion of these contributions is crucial in order to settle the issue of frame dependence.

Another critical point is related to the question of multiloop calculations. For higher loops it seems that one cannot simply use the same tree level transformation rules between the Jordan frame and Einstein frame parametrizations in order to relate and compare the off shell quantum corrected results. Instead, the relevant transformations should be constructed order by order in perturbation theory. This means that one has to find a quantum corrected transformation between the off shell effective actions. However, it does not seem to be clear how to construct such a transformation explicitly. Moreover, even if there might exist such a quantum corrected transformation between different parametrizations, which could be written down explicitly for the specific case under study, such a quantum corrected transformation is of no practical use as long as there is no universal principle of how to obtain this transformation in general. This problem becomes obvious in view of the fact that beside the cosmological 
Jordan 
frame and Einstein frame parametrizations, there are infinitely many other equally valid parametrizations. Similarly, in order to perform the on shell comparison of the $n$-loop corrected effective actions, one should use the ($n-1$)-loop corrected effective equations of motion. In view of this rather complicated iterative strategy it would be desirable to use instead an improved formalism that automatically guarantees the off shell parametrization independence by providing a ``unique'' off shell effective action.

Ultimately, physical results should not depend on the choice of coordinates - or parametrizations in the field theoretical context. Therefore, on the basis of geometrical considerations an explicit construction of a parametrization independent off shell extension of the effective action was proposed and developed in \cite{Vilkovisky}.
By identifying two different parametrizations as two coordinate systems in the configuration space manifold, the off-shell quantum nonequivalence of frames can be traced back to the noncovariant definition of the off shell effective action. If we further identify the transformation between the Jordan frame and the Einstein frame as one particular coordinate transformation in the configuration space of fields \cite{WeProc}, we can apply the ideas of the improved geometrical formalism \cite{Vilkovisky}.
The geometric construction of the effective action in the field theoretical context \cite{Vilkovisky}, offers a new viewpoint on the cosmological debate, Jordan frame vs. Einstein frame, and could also entail important physical consequences.\footnote{We will discuss the physical implications in detail in \cite{WeNew}. The question of equivalence of cosmological solutions in different frames was studied in \cite{Kamenshchik:2013dga, Fre}.
For a recent application of the geometrical formalism to the model of nonminimal Higgs inflation, which was already proposed in \cite{WeProc}, see \cite{Moss:2014nya}. The question of quantum frame dependence could also be interesting for the recently investigated ``quantum tunneling of the universe'' in scalar-tensor theories \cite{WeTunneling,Calcagni:2014xca}. 
}
As has been already emphasized in \cite{WeProc}, from the covariant geometrical point of view, the cosmological debate about a physically preferred frame seems to become meaningless, in the sense that the origin of the discrepancy between the frames at the quantum level is to be found in the mathematical formalism rather than in the physical properties of a specific frame.
Thus, the geometrical viewpoint offers a natural resolution of the controversial discussion about whether the Einstein frame or the Jordan frame should be the ``correct physical frame'': there is no preferred physical frame. Any frame is as good as any other, as long as we work in a covariant formalism.


\section*{ACKNOWLEDGMENTS}

We are grateful to A. O. Barvinsky, C. Kiefer and A. A. Starobinsky for numerous fruitful discussions. A. K. is thankful for helpful discussions with E. Pozdeeva, A. Tronconi, G. Venturi and S. Vernov. C. S. also benefited from stimulating conversations with F. Bezrukov, C. Germani, D. Kaiser, K. Krasnov, M. Shaposhnikov, Yu. Shtanov, J. van der Bij and R. Woodard about this topic. We are grateful to E.~Alvarez, G.~Cognola, S.~ Nojiri and S.~Zerbini for the correspondence on this work.
A. K. acknowledges support by Grant No. 436 RUS 17/3/07 of the German Science Foundation (DFG) and of the University of Cologne for his visits to the University of Cologne. A. K. was partially supported by the RFBR Grant No. 14-02-00894.
C. S. acknowledges support by the Villigst foundation and by the ERC Starting Grant No. 277570-DIGT under which part of this work was done.

\appendix


\section{CONFORMAL TRANSFORMATION}\label{AppA}

Let us consider the following conformal transformation of the metric field $g_{\mu\nu}$ in four space-time dimensions
\begin{align}
g_{\mu\nu}=f\,\hat{g}_{\mu\nu}\,,\label{conformalMetric}
\end{align}
and its inverse
\begin{align}
g^{\mu\nu}=f^{-1}\,\hat{g}^{\mu\nu}\,.
\end{align}
The square root of the determinant transforms as
\begin{align}
\sqrt{g}=f^{2}\,\sqrt{\hat{g}}\,.\label{cofdet}
\end{align}
In order to calculate the transformation of the Riemann tensor, it is necessary to first calculate the behavior of the Christoffel symbol and its first derivative under the conformal transformation (\ref{conformalMetric}),
\begin{align}
\Gamma^{\alpha}_{\mu\nu}=&{}\,\hat{\Gamma}^{\alpha}_{\mu\nu}+\frac{1}{2}\,f^{-1}\Big(\delta_{\mu}^{\alpha}\,f_{,\,\nu}+\delta_{\nu}^{\alpha}f_{,\,\mu}-\hat{g}^{\alpha\gamma}\hat{g}_{\mu\nu}\,f_{,\,\gamma}\Big)\;,\\
\Gamma^{\alpha}_{\mu\nu,\,\beta}=&\hat{\Gamma}^{\alpha}_{\mu\nu,\,\beta}+\frac{1}{2}\,f^{-1}\Big(\delta_{\mu}^{\alpha}\,f_{,\,\nu\mu}+\delta_{\nu}^{\alpha}\,f_{,\mu\beta}-\hat{g}^{\alpha\gamma}_{\;\;\;,\,\beta}\,\hat{g}_{\mu\nu}\,f_{,\,\gamma}
\nonumber\\
&-\hat{g}^{\alpha\gamma}\hat{g}_{\mu\nu,\,\beta}\,f_{,\,\gamma}-\hat{g}^{\alpha\gamma}\hat{g}_{\mu\nu}\,f_{,\,\gamma\beta}\Big)\nonumber\\
&-\frac{1}{2}f^{-2}f_{,\,\beta}\,\Big(
\delta_{\mu}^{\alpha}\,f_{,\,\nu}+\delta_{\nu}^{\alpha}f_{,\,\mu}-\hat{g}^{\alpha\gamma}\hat{g}_{\mu\nu}f_{,\,\gamma}\Big)\,.
\end{align}
In order to facilitate the calculation of the Riemann tensor, Ricci tensor and Ricci scalar, it is convenient to use a Riemannian normal coordinate system for the intermediate calculations in which $\hat{\Gamma}^{\alpha}_{\mu\nu}=0$ but $\hat{\Gamma}^{\alpha}_{\mu\nu,\,\beta}\neq0$.
The final result for the conformal transformation of the Riemann tensor is then found to be
\begin{align}
 R^{\alpha}_{\:\beta\gamma\delta}=&\,\hat{R}^{\alpha}_{\;\beta\gamma\delta}+\frac{1}{4}\,f^{-2}\,\Big(3f_{;\,\gamma}\,f^{;\,\alpha}\,\hat{g}_{\beta\delta}-3\,f_{;\,\delta}\,f^{;\,\alpha}\,\hat{g}_{\beta\gamma}\nonumber\\
&+3\,f_{;\,\beta}\,f_{;\,\delta}\,\delta^{\alpha}_{\gamma}-3\,f_{;\,\beta}f_{;\,\gamma}\,\delta^{\alpha}_{\delta}+f_{;\nu}\,f^{;\;\nu}\,\hat{g}_{\beta\gamma}\delta_{\delta}^{\alpha}\nonumber\\
&-f_{;\nu}\,f^{;\;\nu}\,\hat{g}_{\beta\delta}\,\delta_{\gamma}^{\alpha}\Big)+\frac{1}{2}\,f^{-1}\,\Big(f_{;\,\delta}^{\;\;\alpha}\,\hat{g}_{\beta\gamma}-f_{;\,\gamma}^{\;\;\alpha}\,\hat{g}_{\beta\delta}\nonumber\\
&+f_{;\,\beta\gamma}\,\delta_{\delta}^{\alpha}-f_{;\,\beta\delta}\,\delta_{\gamma}^{\alpha}\Big)\,.
\end{align}
Contracting the first and the third indices the result for the Ricci tensor is
\begin{align}
  R_{\alpha\beta}=&\,\hat{R}_{\alpha\beta}+\frac{3}{2}\,f^{-2}\,f_{;\,\alpha}\,f_{;\,\beta}-f^{-1}\,f_{;\,\alpha\beta}-\frac{1}{2}f^{-1}\,f_{;\,\nu}^{\;\;\nu}\,\hat{g}_{\alpha\beta}\,.
\end{align}
The transformation of the Ricci scalar follows from
\begin{align}
 R=g^{\alpha\beta}R_{\alpha\beta}=f^{-1}\,\hat{R}+\frac{3}{2}\,f^{-3}\,f_{;\nu}\,f^{;\,\nu}-3\,f^{-2}\,f_{;\nu}^{\;\;\nu}\,.\label{ConfRicciScalar}
\end{align}
In the divergent part of the one-loop contributions to the effective action also the structures  $R^{\alpha\beta\gamma\delta}R_{\alpha\beta\gamma\delta}$, $R^{\alpha\beta}R_{\alpha\beta}$ and $R^2$ appear.
By making use of the Gauss-Bonnet identity
\begin{align}
 \frac{\delta}{\delta g_{\mu\nu}}\Bigg[\int\text{d}^4\,x\,\sqrt{g}\,\Bigg\{R^{\alpha\beta\gamma\delta}R_{\alpha\beta\gamma\delta}+4\,R^{\alpha\beta}R_{\alpha\beta}-R^2\Bigg\}\Bigg]=0\,,\label{GaussBonnet}
\end{align}
the structure $R^{\alpha\beta\gamma\delta}R_{\alpha\beta\gamma\delta}$ can be expressed in terms of the structures $R^{\alpha\beta}R_{\alpha\beta}$ and $R^2$.
The transformation laws for these two structures are given by
\begin{align}
 R^{\alpha\beta}R_{\alpha\beta}=&\,f^{-2}\,\hat{R}^{\alpha\beta}\,\hat{R}_{\alpha\beta}-f^{-3}\,\Big(\hat{R}\,f_{;\,\nu}^{\;\;\nu}+2\,\hat{R}^{\alpha\beta}\,f_{;\,\alpha\beta}\Big)\nonumber\\
&+f^{-4}\,\Big(f_{;\,\alpha\beta}\,f^{;\,\alpha\beta}+3\,\hat{R}^{\alpha\beta}\,f_{;\,\alpha}\,f_{;\,\beta}+2\,f_{;\,\nu}^{;\;\nu}\,f_{;\,\mu}^{\;\;\mu}\Big)\nonumber\\
&-3\,f^{-5}\,\Big(\frac{1}{2}\,f_{;\,\nu}\,f^{;\,\nu}\,f_{;\,\mu}^{\;\;\mu}+f^{;\,\alpha\beta}\,f_{;\,\alpha}\,f_{;\,\beta}\Big)\nonumber\\
&+\frac{9}{4}\,f^{-6}\,f^{;\,\alpha}\,f_{;\,\alpha}\,f^{;\,\beta}\,f_{;\,\beta}\,\label{ConfRicciTensorSquare},\\
 R^2=&\,f^{-2}\,\hat{R}^2-6\,f^{-3}\,\hat{R}\,f_{;\,\nu}^{\;\;\nu}+f^{-4}\,\Big(3\,\hat{R}\,f_{;\nu}\,f^{;\nu}\nonumber\\
&+9\,f_{;\,\nu}^{\;\;\nu}\,f_{;\,\mu}^{\;\;\mu}\Big)-9\,f^{-5}\,f_{;\,\nu}\,f^{;\,\nu}\,f_{;\,\mu}^{\;\;\mu}\nonumber\\
&+\frac{9}{4}\,f^{-6}\,f_{;\,\nu}\,f^{;\,\nu}\,f_{;\,\mu}f^{;\,\mu}\label{ConfRicciScalarSquare}\,.
\end{align}


\section{JORDAN FRAME COEFFICIENTS}\label{AppB}

In this appendix we present the explicit form of the coefficients for the Jordan frame effective action (\ref{JFEFFAction})
\begin{widetext}
\begin{align}
U^{\rm{J}}_{1\text{-loop}}=&\;V
   \left[s^2 \left(\frac{4 \left(U'\right)^4}{U^3}-\frac{2
   \left(U'\right)^2 U''}{U^2}\right)-\frac{7\, s
   \left(U'\right)^2}{3\, U^2}-\frac{13}{3\, U}\right]+V' \left[s^2 \left(\frac{4\, U' U''}{U}-\frac{8
   \left(U'\right)^3}{U^2}\right)+\frac{8\, s\, U'}{3\, U}\right]\nonumber\\
&+V'' \left[s^2
   \left(\frac{2
   \left(U'\right)^2}{U}-U''\right)-\frac{s}{6}\right]\,,
    \label{U1loop}
\end{align}
\begin{align}
G^{\rm{J}}_{1\text{-loop}}=&V \left[
s^3 \left(-\frac{18 G' \left(U'\right)^3 U''}{U^3}-\frac{9 G'\left(U'\right)^5}{U^4}-\frac{5 \left(G'\right)^2 \left(U'\right)^2}{2 U^2}-\frac{90 \left(U'\right)^6 U''}{U^5}-\frac{18 \left(U'\right)^4\left(U''\right)^2}{U^4}+\frac{99 \left(U'\right)^8}{2 U^6}
\right)\right.\nonumber\\
&+s^2 \left(
\frac{G'' \left(U'\right)^2}{U^2}+\frac{5 G' U' U''}{2U^2}+\frac{11 G' \left(U'\right)^3}{2 U^3}+\frac{147 \left(U'\right)^4 U''}{2 U^4}+\frac{9 \left(U'\right)^2 \left(U''\right)^2}{U^3}-\frac{105\left(U'\right)^6}{2 U^5}
\right)\nonumber\\
&+\left.s \left(
-\frac{G' U'}{2 U^2}-\frac{17 \left(U'\right)^2 U''}{U^3}+\frac{37
\left(U'\right)^4}{2U^4}\right)+\frac{G}{U^2}+\frac{17 \left(U'\right)^2}{U^3}+\frac{3 U''}{U^2}
\right]+V' \left[
s^3 \left(
\frac{36 G' \left(U'\right)^2 U''}{U^2}\right.\right.\nonumber\\
&\left.+\frac{18 G' \left(U'\right)^4}{U^3}+\frac{5 \left(G'\right)^2U'}{U}+\frac{180 \left(U'\right)^5 U''}{U^4}+\frac{36 \left(U'\right)^3 \left(U''\right)^2}{U^3}-\frac{99 \left(U'\right)^7}{U^5}
\right)+s^2\left(
-\frac{2 G'' U'}{U}\right.\nonumber\\
&\left.-\frac{5 G' U''}{2 U}-\frac{13 G' \left(U'\right)^2}{2 U^2}-\frac{195 \left(U'\right)^3 U''}{2 U^3}-\frac{15U' \left(U''\right)^2}{U^2}+\frac{159 \left(U'\right)^5}{2 U^4}
\right)+s \left(
\frac{G'}{2 U}+\frac{14 U' U''}{U^2}\right.\nonumber\\
&\left.\left.-\frac{47\left(U'\right)^3}{2 U^3}
\right)
-\frac{3 U'}{U^2}
\right]+V'' \left[
s^3 \left(
-\frac{9 G' U'U''}{U}-\frac{9 G' \left(U'\right)^3}{2 U^2}-\frac{5 \left(G'\right)^2}{4}-\frac{45 \left(U'\right)^4 U''}{U^3}\right.\right.\nonumber\\
&\left.-\frac{9 \left(U'\right)^2\left(U''\right)^2}{U^2}+\frac{99 \left(U'\right)^6}{4 U^4}
\right)+s^2 \left(
-\frac{G' U'}{U}+\frac{G''}{2}+\frac{12 \left(U'\right)^2U''}{U^2}-\frac{27 \left(U'\right)^4}{U^3}+\frac{3 \left(U''\right)^2}{U}
\right)\nonumber\\
&+\left.s \left(\frac{11\left(U'\right)^2}{U^2}-\frac{U''}{U}
\right)
-\frac{2}{U}
\right]
+V^{'''} \left[
s^2 \left(
\frac{G'}{2}+\frac{9 \left(U'\right)^3}{2U^2}
\right)
-\frac{s U'}{U}
\right]\,,\label{G1loop}
\end{align}
\begin{align}
\alpha_{1}^{\rm{J}}=\frac{43}{60}+2\,s\,\frac{ \left(U'\right)^2}{U}\label{Alpha1JF1L}\,,
\end{align}
\begin{align}
\alpha_{2}^{\rm{J}}={}&\frac{1}{40}+s^2 \left(-\frac{2 \left(U'\right)^2 U''}{U}+\frac{2
   \left(U'\right)^4}{U^2}+\frac{\left(U''\right)^2}{2}\right)+s \left(\frac{U''}{6}-\frac{4 \left(U'\right)^2}{3
   U}\right)\label{Alpha2JF1L}\,,
\end{align}
\begin{align}
\alpha_{3}^{\rm{J}}={}&s^2 \left(\frac{G' U''}{2}-\frac{G'
   \left(U'\right)^2}{U}+\frac{9 \left(U'\right)^3 U''}{2
   U^2}-\frac{9 \left(U'\right)^5}{U^3}\right)+s\left(\frac{G'}{12}-\frac{U' U''}{U}+\frac{19
   \left(U'\right)^3}{4 U^2}\right)-\frac{3 U'}{U}\,,
\end{align}
\begin{align}
\alpha_{4}^{\rm{J}}={}&s^3 \left(
-\frac{15 G' \left(U'\right)^3 U''}{2
   U^2}+\frac{3 G' U'
   \left(U''\right)^2}{U}+\frac{1}{4} \left(G'\right)^2
   U''+\frac{3 G'
   \left(U'\right)^5}{U^3}-\frac{\left(G'\right)^2
   \left(U'\right)^2}{2 U}+\frac{81 \left(U'\right)^6 U''}{4
   U^4}\nonumber\right.\\
&\quad{}\left.-\frac{27 \left(U'\right)^4
   \left(U''\right)^2}{U^3}+\frac{9 \left(U'\right)^2
   \left(U''\right)^3}{U^2}-\frac{9 \left(U'\right)^8}{2
   U^5}\right)
   +s^2 \left(
   \frac{G' U'
   U''}{U}+\frac{5 G' \left(U'\right)^3}{4
   U^2}+\frac{\left(G'\right)^2}{24}-\frac{12 \left(U'\right)^4
   U''}{U^3}\nonumber\right.\\
&\quad{}\left.   +\frac{18 \left(U'\right)^2
   \left(U''\right)^2}{U^2}+\frac{15 \left(U'\right)^6}{8
   U^4}-\frac{3 \left(U''\right)^3}{U}\right)+s
   \left(-\frac{G' U'}{2 U}-\frac{7 \left(U'\right)^2
   U''}{4 U^2}+\frac{\left(U'\right)^4}{2
   U^3}-\frac{\left(U''\right)^2}{U}\right)-\frac{G}{3
   U}\nonumber\\
&\quad{}   -\frac{19 \left(U'\right)^2}{4 U^2}-\frac{2
   U''}{U}\,,
\end{align}
\begin{align}
\alpha_{5}^{\rm{J}}={}&s^2 \left(-\frac{G' U' U''}{U}-\frac{3 G'
   \left(U'\right)^3}{U^2}-\frac{15 \left(U'\right)^4
   U''}{U^3}-\frac{6 \left(U'\right)^2
   \left(U''\right)^2}{U^2}+\frac{9
   \left(U'\right)^6}{U^4}\right)\nonumber\\
   &{}+s \left(\frac{G'
   U'}{U}+\frac{10 \left(U'\right)^2
   U''}{U^2}-\frac{5 \left(U'\right)^4}{U^3}+\frac{2
   \left(U''\right)^2}{U}\right)-\frac{\left(U'\right)^2}{U^
   2}-\frac{2 U''}{U}\,,
\end{align}
\begin{align}
\alpha_{6}^{\rm{J}}=s^2 \left(\frac{9 G' \left(U'\right)^3}{4
   U^2}+\frac{\left(G'\right)^2}{8}+\frac{81 \left(U'\right)^6}{8
   U^4}\right)+s \left(-\frac{G' U'}{2 U}-\frac{15
   \left(U'\right)^4}{2 U^3}\right)+\frac{27 \left(U'\right)^2}{4
   U^2}-\frac{U''}{U}\,,
\end{align}
\begin{align}
\alpha_{7}^{\rm{J}}={}&s^3 \left(\frac{9 G' \left(U'\right)^4
   U''}{U^3}+\frac{9 G' \left(U'\right)^2
   \left(U''\right)^2}{2 U^2}+\frac{3 \left(G'\right)^2 U'
   U''}{2 U}-\frac{45 G' \left(U'\right)^6}{8
   U^4}+\frac{3 \left(G'\right)^2 \left(U'\right)^3}{8
   U^2}+\frac{\left(G'\right)^3}{8}\nonumber\right.\\
&\quad{}\left.-\frac{81 \left(U'\right)^7
   U''}{2 U^5}+\frac{81 \left(U'\right)^5
   \left(U''\right)^2}{2 U^4}+\frac{81 \left(U'\right)^9}{8
   U^6}\right)+s^2 \left(-\frac{3 G' \left(U'\right)^2
   U''}{4 U^2}-\frac{3 G' \left(U''\right)^2}{2
   U}-\frac{3 G' \left(U'\right)^4}{2
   U^3}\right.\nonumber\\
&\quad{}\left.-\frac{\left(G'\right)^2 U'}{4 U}+\frac{153
   \left(U'\right)^5 U''}{4 U^4}-\frac{63 \left(U'\right)^3
   \left(U''\right)^2}{2 U^3}-\frac{45 \left(U'\right)^7}{4
   U^5}\right)+s \left(\frac{G' U''}{2
   U}+\frac{G' \left(U'\right)^2}{U^2}-\frac{9
   \left(U'\right)^3 U''}{U^3}\right.\nonumber\\
&\quad{}\left.+\frac{15 U'
   \left(U''\right)^2}{2 U^2}+\frac{9 \left(U'\right)^5}{2
   U^4}\right)+\frac{G'}{2 U}-\frac{9 G U'}{2
   U^2}+\frac{15 U' U''}{U^2}-\frac{15
   \left(U'\right)^3}{2 U^3}-\frac{U^{'''}}{U}\,,
\end{align}
\begin{align}
\alpha_{8}^{\rm{J}}={}&\left(\frac{81 \left(U'\right)^{12}}{32 U^8}-\frac{81 U''
   \left(U'\right)^{10}}{4 U^7}-\frac{27 G'
   \left(U'\right)^9}{8 U^6}+\frac{243 \left(U''\right)^2
   \left(U'\right)^8}{4 U^6}+\frac{81 G' U''
   \left(U'\right)^7}{4 U^5}-\frac{81 \left(U''\right)^3
   \left(U'\right)^6}{U^5}\nonumber\right.\\
&\quad{}\left.+\frac{27 \left(G'\right)^2
   \left(U'\right)^6}{16 U^4}-\frac{81 G' \left(U''\right)^2
   \left(U'\right)^5}{2 U^4}+\frac{81 \left(U''\right)^4
   \left(U'\right)^4}{2 U^4}-\frac{27 \left(G'\right)^2 U''
   \left(U'\right)^4}{4 U^3}-\frac{3 \left(G'\right)^3
   \left(U'\right)^3}{8 U^2}\right.\nonumber\\
&\quad{}\left.+\frac{27 G' \left(U''\right)^3
   \left(U'\right)^3}{U^3}+\frac{27 \left(G'\right)^2
   \left(U''\right)^2 \left(U'\right)^2}{4 U^2}+\frac{3
   \left(G'\right)^3 U'' U'}{4
   U}+\frac{\left(G'\right)^4}{32}\right) s^4+\left(\frac{9
   \left(U'\right)^{10}}{4 U^7}+\frac{135 U''
   \left(U'\right)^8}{8 U^6}\right.\nonumber\\
&\quad{}\left.-\frac{3 G' \left(U'\right)^7}{2
   U^5}-\frac{405 \left(U''\right)^2 \left(U'\right)^6}{4
   U^5}-\frac{57 G' U'' \left(U'\right)^5}{4
   U^4}+\frac{261 \left(U''\right)^3 \left(U'\right)^4}{2
   U^4}+\frac{\left(G'\right)^2 \left(U'\right)^4}{4
   U^3}+\frac{39 G' \left(U''\right)^2
   \left(U'\right)^3}{U^3}\right.\nonumber\\
&\quad{}\left.-\frac{27 \left(U''\right)^4
   \left(U'\right)^2}{U^3}+\frac{23 \left(G'\right)^2 U''
   \left(U'\right)^2}{8 U^2}-\frac{9 G' \left(U''\right)^3
   U'}{U^2}-\frac{3 \left(G'\right)^2 \left(U''\right)^2}{4
   U}\right) s^3+\left(-\frac{45 \left(U'\right)^8}{4
   U^6}\right.\nonumber\\
&\quad{}\left.+\frac{141 U'' \left(U'\right)^6}{4 U^5}+\frac{29
   G' \left(U'\right)^5}{4 U^4}-\frac{3 U^{'''}
   \left(U'\right)^5}{2 U^4}-\frac{135 \left(U''\right)^2
   \left(U'\right)^4}{8 U^4}-\frac{G'' \left(U'\right)^4}{2
   U^3}-\frac{67 G' U'' \left(U'\right)^3}{4
   U^3}\nonumber\right.\\
&\quad{}\left.-\frac{6 U'' U^{'''}
   \left(U'\right)^3}{U^3}-\frac{39 \left(U''\right)^3
   \left(U'\right)^2}{2 U^3}-\frac{\left(G'\right)^2
   \left(U'\right)^2}{U^2}-\frac{G'' U''
   \left(U'\right)^2}{2 U^2}-\frac{G' U^{'''}
   \left(U'\right)^2}{2 U^2}+\frac{4 G' \left(U''\right)^2
   U'}{U^2}\nonumber\right.\\
&\quad{}\left.+\frac{9 \left(U''\right)^4}{2
   U^2}+\frac{\left(G'\right)^2 U''}{2 U}\right)
   s^2+\left(\frac{25 \left(U'\right)^6}{2 U^5}-\frac{83 U''
   \left(U'\right)^4}{2 U^4}-\frac{17 G'
   \left(U'\right)^3}{4 U^3}+\frac{9 U^{'''}
   \left(U'\right)^3}{2 U^3}+\frac{133 \left(U''\right)^2
   \left(U'\right)^2}{4 U^3}\nonumber\right.\\
&\quad{}\left.+\frac{G'' \left(U'\right)^2}{2
   U^2}+\frac{7 G' U'' U'}{U^2}-\frac{U''
   U^{'''} U'}{U^2}-\frac{\left(U''\right)^3}{2
   U^2}+\frac{\left(G'\right)^2}{4 U}-\frac{G'
   U^{'''}}{2 U}\right) s+\frac{145 \left(U'\right)^4}{16
   U^4}+\frac{27 G \left(U'\right)^2}{4 U^3}\nonumber\\
&\quad{}+\frac{7
   \left(U''\right)^2}{2 U^2}-\frac{3 G'
   U'}{U^2}+\frac{3 \left(U'\right)^2
   U''}{U^3}+\frac{2 G U''}{U^2}-\frac{3 U'
   U^{'''}}{2 U^2}+\frac{5 G^2}{4 U^2}\,.
\end{align}
\end{widetext}


\section{EINSTEIN FRAME COEFFICIENTS}\label{AppC}

In this appendix we present the explicit form of the coefficients in (\ref{EffActionEF}), expressed in terms of the Jordan frame parametrization. We start with the two most important remaining structures 
\begin{widetext}
\begin{align}
U_{1-\rm{loop}}^{\rm{E}}={}&V \left[s^2 \left(-\frac{G'
   U'}{6 U}-\frac{\left(U'\right)^2
   U''}{U^2}+\frac{\left(U'\right)^4}{2
   U^3}\right)+s \left(\frac{U''}{3 U}-\frac{5
   \left(U'\right)^2}{6 U^2}\right)-\frac{13}{3
   U}\right]\nonumber\\
&\quad{}+V' \left[s^2 \left(\frac{G'}{12}+\frac{U' U''}{2
   U}-\frac{\left(U'\right)^3}{4 U^2}\right)+\frac{7 s
   U'}{12 U}\right]
-s\,\frac{V''}{6}\,,
\end{align}
\begin{align}
G_{1-\rm{loop}}^{\rm{E}}&={}
V'' \left[s^2 \left(\frac{3 G' U'}{4 U}+\frac{9
   \left(U'\right)^2 U''}{2 U^2}-\frac{9
   \left(U'\right)^4}{4 U^3}\right)+\frac{2 s
   \left(U'\right)^2}{U^2}-\frac{2}{U}\right]+V'\,\,\left[
s^3 \left(-\frac{6 G' \left(U'\right)^2
   U''}{U^2}\right.\right.\nonumber\\
&\quad{}\left.\left.+\frac{3 G'
   \left(U'\right)^4}{U^3}-\frac{\left(G'\right)^2 U'}{2
   U}+\frac{18 \left(U'\right)^5 U''}{U^4}-\frac{18
   \left(U'\right)^3 \left(U''\right)^2}{U^3}-\frac{9
   \left(U'\right)^7}{2 U^5}\right)+s^2 \left(\frac{G''
   U'}{4 U}-\frac{19 G' \left(U'\right)^2}{8
   U^2}\right.\right.\nonumber\\
&\quad{}\left.\left.-\frac{18 \left(U'\right)^3 U''}{U^3}+\frac{3
   U' \left(U''\right)^2}{2 U^2}+\frac{69
   \left(U'\right)^5}{8 U^4}+\frac{3 U^{(3)}
   \left(U'\right)^2}{2 U^2}\right)+s\left(\frac{G'}{U}+\frac{35 U' U''}{4
   U^2}-\frac{65 \left(U'\right)^3}{8 U^3}\right)-\frac{6
   U'}{U^2}\right]\nonumber\\
&+V\,\left[
s^3 \left(\frac{12 G'
   \left(U'\right)^3 U''}{U^3}-\frac{6 G'
   \left(U'\right)^5}{U^4}+\frac{\left(G'\right)^2
   \left(U'\right)^2}{U^2}-\frac{36 \left(U'\right)^6
   U''}{U^5}+\frac{36 \left(U'\right)^4
   \left(U''\right)^2}{U^4}+\frac{9
   \left(U'\right)^8}{U^6}\right)\right.\nonumber\\
&\quad{}+s^2 \left(-\frac{G''
   \left(U'\right)^2}{2 U^2}-\frac{3 G' U' U''}{2
   U^2}+\frac{13 G' \left(U'\right)^3}{4 U^3}+\frac{63
   \left(U'\right)^4 U''}{2 U^4}-\frac{12 \left(U'\right)^2
   \left(U''\right)^2}{U^3}-\frac{51 \left(U'\right)^6}{4
   U^5}\right.\nonumber\\
&\quad{}\left.-\frac{3 U^{'''}
   \left(U'\right)^3}{U^3}\right)+s \left.\left(-\frac{2 G'
   U'}{U^2}-\frac{37 \left(U'\right)^2 U''}{2
   U^3}+\frac{49 \left(U'\right)^4}{4 U^4}+\frac{U^{'''}
   U'}{U^2}\right)+\frac{G}{U^2}+\frac{25
   \left(U'\right)^2}{2 U^3}+\frac{4
   U''}{U^2}\right]\nonumber\\
&\quad{}-\frac{V''' s U'}{2 U}\,.
\end{align}
\end{widetext}
The results for these structures coincide with those derived in \cite{we-renorm}.
The coefficients of the remaining structures are given by
\begin{align}
\alpha_{1}^{\rm{E}}=\frac{43}{60}\,,
\end{align}
\begin{align}
\alpha_{2}^{\rm{E}}=\frac{1}{40}\,,
\end{align}
\begin{align}
\alpha_{3}^{\rm{E}}=-\frac{19}{12}\,\frac{U'}{U}\,,
\end{align}
\begin{align}
\alpha_{4}^{\rm{E}}=-\frac{G}{3\,U}-\frac{5}{24}\,\frac{(U')^2}{U^2}-\frac{19}{12}\,\frac{U''}{U}\,,
\end{align}
\begin{align}
\alpha_{5}^{\rm{E}}=0\,,
\end{align}
\begin{align}
\alpha_{6}^{\rm{E}}=\frac{19}{8}\,\frac{(U')^2}{U^2}\,,\\\nonumber
\end{align}
\begin{align}
\alpha_{7}^{\rm{E}}=\frac{G\,U'}{U^2}+\frac{5}{8}\,\frac{(U')^3}{U^3}+\frac{19}{4}\,\frac{U'\,U''}{U^2}\,,
\end{align}
\begin{align}
\alpha_{8}^{\rm{E}}={}&\frac{5}{4}\,\frac{G^2}{U^2}+7\,\frac{G\,(U')^2}{U^3}+\frac{331}{32}\,\frac{(U')^4}{U^4}\nonumber\\
&+\frac{G\,U''}{U^2}+\frac{5}{8}\,\frac{(U')^2\,U''}{U^3}+\frac{19}{8}\,\frac{(U'')^2}{U^2}\,.\\\nonumber\\\nonumber
\end{align}


\section{CALCULATION OF THE EINSTEIN FRAME EFFECTIVE ACTION}\label{AppD}

In this appendix, we have dropped the convention to denote a field in the Einstein frame parametrization by a hat in order not to overload the notation.
The result of the divergent part for the one-loop contributions to the effective action can be expressed in terms of the heat Kernel coefficients $a_{2}$ of the gauge-fixed action and $a_{2}^{(Q)}$ of the ghost action
\begin{widetext}
\begin{align}
\Gamma^{(1)}_{\rm{div}}=\underset{\omega\to 2}{\lim}\;\frac{1}{32\,\pi^2\,(2-\omega)}\left[\,\int\text{d}^{2\,\omega}x\;\sqrt{g}\;\text{tr}\,[\hat{a}_{2}]-2\int\text{d}^{2\,\omega}x\;\sqrt{g}\;\text{tr}\,[(a^{(Q)}_{2})_{\mu}^{\;\nu}]\,\right]\,.\label{OneLoopActionA2Final}
\end{align}
\end{widetext}
For a general differential operator of the form
\begin{align}
 \hat{F}:=\hat{\bar{\square}}+\hat{P}-\frac{1}{6}\,\bar{R}\,\hat{I}\,,\label{MinimalFormOperator}
\end{align}
the coincidence limit $x\to x'$ of the corresponding $a_{2}$-coefficient has the universal structure \cite{Bar-Vil}
\begin{widetext}
\begin{align}
\hat{a}_{2}(x,x)={}&\frac{\hat{\mathbf{1}}}{180}\left(\bar{R}_{\alpha\beta\mu\nu}
	 ^{2}-\bar{R}_{\mu\nu}^{2}+\bar{\Box} \bar{R}\right)+\frac{1}{2}\,\hat{P}^2+\frac{1}{12}\,\hat{{\cal R}}_{\mu\nu}^2+
\frac{1}{6}\,\bar{\Box}\hat{P}\;.\label{A2Coefficient}
\end{align}
\end{widetext}
In this condensed notation, the hat signifies that bundle indices $A,\,B\,...$ have been suppressed, i.e. $\hat{P}=P_{A}^{\;\;B}$ etc.
Here, $\hat{{\cal R}}_{\mu\nu}$ is the bundle curvature to be introduced below and $\hat{\mathbf{1}}:=\delta_{A}^{\;\;B}$ is the identity in field space.
The differential operators $\hat{F}$  and $\hat{Q}$ are determined by the second variation of the gauge-fixed action and the compensating ghost action
\begin{align}
 S_{,\,ij}^{\rm{tot}}=\frac{\delta^2\,(S+S_{\rm{gf}})}{\delta\phi^{A}(x)\,\delta\phi^{B}(x')}=F_{AB}(\bar{\nabla}^{x})\delta(x,x')\,.\label{SecVar}
\end{align}
Here, $S$ is the Einstein frame action (\ref{actionEF}), and a bar denotes a background quantity, i.e. $\bar{\nabla}_{\mu}$ denotes the covariant derivative with respect to the background connection.
For the action (\ref{actionEF}), we choose the standard DeDonder gauge condition
\begin{align}
\chi^{\rho}(h_{\mu\nu}):=\bar{\nabla}_{\sigma} h^{\sigma\rho}-\frac{1}{2}\bar{\nabla}^{\rho}h=0\,,\;\label{DeDonderGaugeFixing}
\end{align}
with $h:=\bar{g}^{\mu\nu}h_{\mu\nu}$. The gauge breaking term then has the structure
\begin{align}
 S_{\rm{gb}}=-\frac{1}{2}\int\text{d}^4x\,\sqrt{g}\;\chi^{\mu}\,g_{\mu\nu}\chi^{\nu}\;.
\end{align}
Acting with a gauge transformation on the gauge-fixing condition (\ref{DeDonderGaugeFixing}) gives rise to the ghost operator
\begin{align}
Q^{\rho}_{\sigma}:=\frac{\delta\,(\chi_{\xi})^{\rho}}{\delta\xi^{\sigma}}=\bar{\square}\,\delta^{\rho}_{\sigma}+\bar{R}^{\rho}_{\sigma}\;.\label{GhostOperator}
\end{align}
Here, $\xi^{\mu}$ is the vector field pointing in the direction of the Lie dragging.
The differential operator defined by (\ref{SecVar}) has the formal structure
\begin{align}
F_{AB}= C_{AB}\,\bar{\square}+2\,\Gamma_{AB}^{\sigma}\bar{\nabla}_{\sigma}+W_{AB}\;,\label{FlucOppGeneric}
\end{align}
where the different parts are ordered according to the number of derivatives acting on the perturbations.
The individual parts can be read off from the result of (\ref{SecVar}),
\begin{align}
 C_{AB}  =&\left( \begin{array}{ccc}
\bar{G}^{\alpha\beta\gamma\delta}&\quad &0\\
\quad & \quad & \quad \\
0 &\quad  &1\\
\end{array}\right)\,,\label{MatrixC}\\
\Gamma_{AB}^{\sigma}  =&\left( \begin{array}{ccc}
0&\quad &\bar{G}^{\alpha\beta\mu\sigma}\,\bar{\varphi}_{,\,\mu}\\
\quad & \quad & \quad \\
-\bar{G}^{\alpha\beta\mu\sigma}\,\bar{\varphi}_{,\,\mu} &\quad  &0\\
\end{array}\right)\;,\label{LinearPartSingle}\\
W_{AB}=&\left( \begin{array}{ccc}
\bar{G}^{\alpha\beta\mu\nu}\,\bar{P}^{\gamma\delta}_{\mu\nu}&\quad &-\frac{1}{2}\bar{g}^{\alpha\beta}\, \bar{V}'\\
\quad & \quad & \quad \\
 -2\bar{G}^{\gamma\delta\mu\nu}(\bar{\nabla}_{\nu}\bar{\nabla}_{\mu}\bar{\varphi})-\frac{1}{2}\bar{g}^{\gamma\delta}\,\bar{V}'&\quad&-\,\bar{V}''\\
\end{array}\right)\;.\label{PotentialPartSingle}
\end{align}
We have introduced the abbreviation
\begin{align}
\bar{G}^{\alpha\beta\,\gamma\delta}:={}&\frac{1}{4}\left(\bar{g}^{\alpha\gamma}\bar{g}^{\beta\delta}+\bar{g}^{\alpha\delta}\bar{g}^{\beta\gamma}-\bar{g}^{\alpha\beta}\bar{g}^{\gamma\delta}\right),
\end{align}
and its inverse
\begin{align}
\bar{G}_{\alpha\beta\,\gamma\delta}:={}&\bar{g}_{\alpha\gamma}\bar{g}_{\beta\delta}+\bar{g}_{\alpha\delta}\bar{g}_{\beta\gamma}-\bar{g}_{\alpha\beta}\bar{g}_{\gamma\delta}\,,
\end{align}
as well as the potential contribution
\begin{align}
\bar{P}^{\gamma\delta}_{\mu\nu}:= \bar{K}^{\gamma\delta}_{\mu\nu}+\bar{S}^{\gamma\delta}_{\mu\nu}+\delta^{\gamma\delta}_{\mu\nu}\,\bar{V}\,,
\end{align}
with the curvature contribution
\begin{align}
 \bar{K}^{\gamma\delta}_{\lambda\sigma}:={}&2\,\bar{R}_{(\lambda\;\;\;\sigma)}^{\;\;\;\;\gamma\;\;\;\;\delta}+2\delta^{(\gamma}_{(\lambda}\,\bar{R}^{\delta)}_{\sigma)}-\delta^{\gamma\delta}_{\lambda\sigma}\,\bar{R}-\bar{g}^{\gamma\delta}\,\bar{R}_{\lambda\sigma}\nonumber\\
&-\bar{R}^{\gamma\delta}\,\bar{g}_{\lambda\sigma}+\frac{1}{2}\bar{g}^{\gamma\delta}\bar{g}_{\lambda\sigma}\,\bar{R}\,,
\end{align}
and the scalar field contribution
\begin{align}
 \bar{S}^{\gamma\delta}_{\lambda\sigma}:={}&\frac{1}{2}\,\bar{\varphi}_{,\,\mu}\bar{\varphi}_{,\,\nu}\bar{g}^{\mu\nu}\,\delta^{\gamma\delta}_{\lambda\sigma}-2\,\delta^{(\gamma}_{(\lambda}\bar{\varphi}_{,\,\sigma)}\bar{\varphi}^{,\,\delta)}+\frac{1}{2}\,\bar{g}_{\lambda\sigma}\,\bar{\varphi}^{,\,\gamma}\bar{\varphi}^{,\,\delta}\nonumber\\
{}&+\frac{1}{2}\,\bar{g}^{\gamma\delta}\,\bar{\varphi}_{,\,\lambda}\bar{\varphi}_{,\,\sigma}-\frac{1}{4}\,\bar{g}^{\gamma\delta}\bar{g}_{\lambda\sigma}\bar{\varphi}^{,\,\nu}\bar{\varphi}_{,\,\nu}\;.
\end{align}
The operator (\ref{FlucOppGeneric}) has to be brought into its minimal form (\ref{MinimalFormOperator}) suitable for the application of formula (\ref{A2Coefficient}).
This can be accomplished in two steps. First we ``canonicalize'' the D'Alembertian part by multiplying the operator (\ref{FlucOppGeneric}) with the inverse $(C^{-1})^{AB}$ of (\ref{MatrixC}). This corresponds to a redefinition of the perturbation field $\delta\phi^{i}$. The resulting Jacobian in the path integral leads to a factor $\propto\,\delta(0)$ that exactly cancels the volume divergence $\propto\delta(0)$ that was suppressed in the definition (\ref{OneLoopActionA2Final}) for the one-loop approximation \cite{Fradkin}. 
 The operator (\ref{FlucOppGeneric}) then becomes
 \begin{align}
 \tilde{F}_{A}^{\;\;B}=\delta_{A}^{\;\;B}\,\bar{\Box}+2\,\tilde{\Gamma}^{\mu\;B}_{\;A}\,\bar{\nabla}_{\mu}+\tilde{W}_{A}^{\;\;B}\,.\label{NormFlucOp}
 \end{align}
 The second step consists in removing the part of (\ref{NormFlucOp}) which is linear in the derivative. This can be achieved by a redefinition of the covariant derivative
\begin{align}
 \hat{\bar{\nabla}}^{\mu}\to\hat{{\cal D}}^{\mu}:=\hat{\bar{\nabla}}^{\mu}+\hat{\Gamma}^{\mu}\,.\label{ReDefCovDer}
\end{align}
The redefinition, in turn, leads to a modification of the potential $\tilde{W}$ in (\ref{NormFlucOp}) as well as to a modified commutator (bundle) curvature that is now defined with respect to the new covariant derivative $\hat{D}_{\mu}$,
\begin{align}
 [\hat{{\cal D}}_{\mu},\,\hat{{\cal D}}_{\nu}]\,\phi=\hat{{\cal R}}_{\mu\nu}\phi=({\cal R}_{\mu\nu})_{A}^{\;B}\phi_{B}\;.
\end{align}
With these modifications and by absorbing a factor of $1/6\,\bar{R}\,\delta_{A}^{\;\;B}$ into the definition of the potential part, we can bring the operator (\ref{FlucOppGeneric}) into its minimal form (\ref{MinimalFormOperator}).
By using (\ref{ReDefCovDer}), we can express this operator again in terms of the original background derivative $\bar{\nabla}_{\mu}$ and we find the following result for the  potential part and the commutator curvature
\begin{align}
 \hat{P}={}& P_{A}^{B}:=W_{A}^{B}+\frac{1}{6}\,\bar{R}\,\delta^{B}_{A}-(\bar{\nabla}_{\mu}\Gamma_{A}^{\:B\:\mu})\nonumber\\
 &\quad\quad\quad\quad-\bar{g}_{\mu\nu}\Gamma_{A}^{\:B\:\mu}\Gamma_{B}^{\:C\:\nu}\;,\label{PotTermDef}\\\nonumber\\
 \hat{\mathcal{R}}_{\mu\nu}={}&\mathcal{R}_{A\:\:\,\mu\nu}^{\:\:B}:=
\mathcal{R}^{0\:B}_{\:A\:\:\,\mu\nu}+2\,\bar{\nabla}_{[\mu}\bar{g}_{\nu]\lambda}\,\Gamma_{A}^{\;B\:\,\lambda}\nonumber\\
&\quad\quad\quad\quad+2\,\bar{g}_{\sigma[\mu}\bar{g}_{\nu]\lambda}\,\Gamma^{\;C\:\,\sigma}_{A}\,\Gamma^{\;B\:\,\lambda}_{C}\;.\label{FieldCurvatureDef}
\end{align}
Here, the commutator curvature $\hat{{\cal R}}^{0}_{\mu\nu}$ with respect to the original background derivative $\bar{\nabla}_{\mu}$ is defined by
\begin{align}
 [\hat{\bar{\nabla}}_{\mu},\,\hat{\bar{\nabla}}_{\nu}]\,\phi=\hat{{\cal R}}^{0}_{\mu\nu}\,\phi\;.\label{BackgroundCuravtureDef}
\end{align}
Using (\ref{MatrixC})--(\ref{PotentialPartSingle}), the explicit form of the connection part in (\ref{NormFlucOp}) is given by
\begin{align}
 \hat{\Gamma}^{\epsilon}:=&\,\tilde{\Gamma}_{B}^{A\;\epsilon}:=(C^{-1})^{\:AC}\,\Gamma_{CB}^{\;\epsilon}\nonumber\\
=&\left( \begin{array}{ccc}
0&\quad &\delta^{\mu\epsilon}_{\alpha\beta}\,\bar{\varphi}_{,\,\mu}\\
\quad & \quad & \quad \\
-\bar{G}^{\gamma\delta\mu\epsilon}\,\bar{\varphi}_{,\,\mu}  &\quad  &0\\
\end{array}\right)\,,\label{ReDefGamma}
\end{align}
and the potential part by
\begin{align}
 \hat{W}:=&\,\tilde{W}_{B}^{A}:=(C^{-1})^{\:AC}\,W_{CB}\nonumber\\
=&\left( \begin{array}{ccc}
\bar{P}^{\gamma\delta}_{\alpha\beta}&\quad &\bar{g}_{\alpha\beta}\,\bar{V}'\\
\quad & \quad & \quad \\
 -2\bar{G}^{\gamma\delta\mu\nu}(\bar{\nabla}_{\nu}\bar{\nabla}_{\mu}\bar{\varphi})-\frac{1}{2}\bar{g}^{\gamma\delta}\,\bar{V}'&\quad&-\bar{V}''\\
\end{array}\right)\label{ReDefPotential}\;.
\end{align}
Substituting (\ref{ReDefGamma})--(\ref{ReDefPotential})) into (\ref{PotTermDef})--(\ref{FieldCurvatureDef}), we find the explicit expressions
\begin{widetext}
\begin{align}
 \hat{P}={}& P_{A}^{B}=W_{A}^{B}+\frac{1}{6}\,\bar{R}\,\delta^{B}_{A}-(\bar{\nabla}_{\mu}\Gamma_{A}^{\:B\:\mu})-\bar{g}_{\rho\epsilon}\Gamma_{A}^{\:C\,\epsilon}\Gamma_{C}^{\:B\,\rho}\nonumber\\\nonumber\\
={}&\left( \begin{array}{ccc}
\bar{P}^{\gamma\delta}_{\alpha\beta}+\frac{1}{6}\,\bar{R}\,\delta^{\gamma\delta}_{\alpha\beta}+\bar{g}_{\epsilon\rho}\,\delta^{\mu\epsilon}_{\alpha\beta}\,\bar{G}^{\gamma\delta\nu\rho}\,\bar{\varphi}_{,\,\mu}\bar{\varphi}_{,\nu}&\quad &\bar{g}_{\alpha\beta}\, \bar{V}'-(\bar{\nabla}_{\alpha}\bar{\nabla}_{\beta}\bar{\varphi})\\
\quad & \quad & \quad \\
 -\bar{G}^{\gamma\delta\mu\nu}(\bar{\nabla}_{\nu}\bar{\nabla}_{\mu}\bar{\varphi})-\frac{1}{2}\bar{g}^{\gamma\delta}\,\bar{V}' &\quad&-\,\bar{V}''+\frac{1}{6}\bar{R}+(\bar{\varphi}_{,\,\nu}\bar{\varphi}^{,\,\nu})\\
\end{array}\right)\;,\label{FinalPot}\\\nonumber\\
 \hat{\mathcal{R}}_{\mu\nu}={}&\,\mathcal{R}_{A\:\:\,\mu\nu}^{\:\:B}=
\mathcal{R}^{0\:B}_{\:A\:\:\,\mu\nu}+2\,\bar{\nabla}_{[\mu}\bar{g}_{\nu]\epsilon}\,\Gamma_{A}^{B\:\,\epsilon}+2\,\bar{g}_{\epsilon[\mu}\bar{g}_{\nu]\rho}\,\Gamma^{C\:\,\epsilon}_{A}\,\Gamma^{B\:\,\rho}_{C}\nonumber\\\nonumber\\
={}&\left( \begin{array}{ccc}
-2\delta_{(\alpha}^{(\gamma}\bar{R}^{\delta)}_{\beta)\mu\nu}-2\bar{g}_{\epsilon[\mu}\bar{g}_{\nu]\rho}\delta^{\tau\epsilon}_{\alpha\beta}\bar{G}^{\gamma\delta\eta\rho}\,\bar{\varphi}_{,\,\tau}\,\bar{\varphi}_{,\,\eta}&\quad&2\bar{\nabla}_{[\mu}\bar{g}_{\nu]\epsilon}\delta_{\alpha\beta}^{\lambda\epsilon}\bar{\varphi}_{;\,\lambda}\\
\quad & \quad & \quad \\
 -2\bar{\nabla}_{[\mu}\bar{g}_{\nu]\epsilon}\,\bar{G}^{\gamma\delta\lambda\epsilon}\bar{\varphi}_{;\,\lambda} &\quad  &0\\
\end{array}\right)\;.\label{FinalCurv}
\end{align}
\end{widetext}
Since a trace appears in the final expression (\ref{OneLoopActionA2Final}), we have to calculate the trace of the square of these quantities:
\begin{align}
\text{tr}(\hat{P}\hat{P})\label{TrPotSqr}
={}&3\bar{R}_{\alpha\beta\gamma\delta}\bar{R}^{\alpha\beta\gamma\delta}-6\,\bar{R}_{\mu\nu}\bar{R}^{\mu\nu}+\frac{119}{36}\,\bar{R}^2\nonumber\\
&-\frac{5}{6}\bar{R}(\bar{\varphi}_{,\,\nu}\bar{\varphi}^{,\,\nu})+\frac{11}{4}\,(\bar{\varphi}_{,\,\nu}\bar{\varphi}^{,\,\nu})^2+10\,\bar{V}^2\nonumber\\
&-\frac{26}{3}\,\bar{R}\,\bar{V}-\frac{1}{3}\,\bar{R}\,\bar{V}''-4\,\bar{V}'^2+\bar{V}''^2\nonumber\\
&+2\bar{V}'\,\bar{\square}\bar{\varphi}+2\,\bar{V}\,(\bar{\varphi}_{,\,\nu}\bar{\varphi}^{,\,\nu})-2\bar{V}''\,(\bar{\varphi}_{,\,\nu}\bar{\varphi}^{,\,\nu})\nonumber\\
&+\bar{\varphi}_{;\,\mu\nu}\,\bar{\varphi}^{;\,\mu\nu}-\frac{1}{2}\,(\bar{\square}\bar{\varphi})^2\,,\\ \nonumber\\
 \text{tr}(\hat{\mathcal{R}}_{\mu\nu}\hat{\mathcal{R}}^{\mu\nu})=&-6\bar{R}_{\alpha\beta\gamma\delta}\bar{R}^{\alpha\beta\gamma\delta}+\bar{R}(\bar{\varphi}_{,\,\nu}\bar{\varphi}^{,\,\nu})\nonumber\\
 &+2\bar{R}_{\mu\nu}(\bar{\varphi}^{,\,\mu}\bar{\varphi}^{,\nu})-\frac{3}{2}\,(\bar{\varphi}_{,\,\nu}\bar{\varphi}^{,\,\nu})^2\nonumber\\
&-4\bar{\varphi}_{;\,\mu\nu}\bar{\varphi}^{;\,\mu\nu}+(\bar{\square}\bar{\varphi})^2\,.
\end{align}
The trace over the identity $\text{tr}(\hat{\mathbf{1}})=11$ is composed of the ten degrees of freedom contained in the metric field $g_{\mu\nu}$ plus one degree of freedom from the scalar field $\varphi$.

It remains to calculate the ghost contribution. Following the same steps as for the operator (\ref{FlucOppGeneric}), we obtain
\begin{align}
\text{tr}^{(Q)}(\mathbf{1})={}&4\,,\\
\text{tr}^{(Q)}(\hat{P}^2)={}&\bar{R}_{\alpha\beta}\bar{R}^{\alpha\beta}+\frac{4}{9}\bar{R}^2\,,\label{GhostPot}\\
\text{tr}^{(Q)}(\hat{\mathcal{R}}_{\mu\nu}\hat{\mathcal{R}}^{\mu\nu})={}&-\bar{R}_{\alpha\beta\gamma\delta}\bar{R}^{\alpha\beta\gamma\delta}\label{FieldCurvGhost}\;.
\end{align}
Using (\ref{OneLoopActionA2Final}), inserting the results (\ref{TrPotSqr})--(\ref{FieldCurvGhost}) in the formulas for the $a_{2}$-coefficient (\ref{A2Coefficient}), performing an integration by parts of the structures $(\bar{\nabla}_{\mu}\bar{\nabla}_{\nu}\bar{\varphi})(\bar{\nabla}^{\mu}\bar{\nabla}^{\nu}\bar{\varphi})$, $\bar{V}'\bar{\square}\bar{\varphi}$ and making use of the topological Gauss-Bonnet identity (\ref{GaussBonnet}), we obtain the final result for the effective action in the Einstein frame parametrization (\ref{EffActionEF}).


\begin{thebibliography}{99}


\bibitem{Spok}
B.~L.~Spokoiny,
Inflation and generation of perturbations in broken symmetric theory of
gravity,
Phys. Lett. {\bf 147B}, 39 (1984).

\bibitem{Fakir}
R.~Fakir and W.~G.~Unruh,
Improvement on cosmological chaotic inflation through nonminimal
coupling,
Phys.~Rev.~D {\bf 41}, 1783 (1990).

\bibitem{Salop}
D.~S.~Salopek, J.~R.~Bond, and J.~M.~Bardeen,
Designing density fluctuation spectra in inflation,
Phys.~Rev.~D {\bf 40}, 1753 (1989).

\bibitem{Komatsu}
E.~Komatsu and T.~Futamase,
Complete constraints on a nonminimally coupled chaotic inflationary
scenario from the cosmic microwave background,
Phys.~Rev.~D {\bf 59}, 064029 (1999).

\bibitem{Star-nonmin}
B.~Boisseau, G.~Esposito-Far\`ese, D.~Polarski and A.~A.~Starobinsky, Reconstruction of a Scalar-Tensor Theory of Gravity in an Accelerated Universe,
Phys.~Rev.~Lett.~{\bf 85}, 2235 (2000).

\bibitem{we}
A.~O.~Barvinsky and A.~Yu.~Kamenshchik,
One-loop quantum cosmology: the normalizability of the Hartle-Hawking wave
function and the probability of inflation,
Classical Quantum Gravity~{\bf 7}, L181 (1990).

\bibitem{we1}
A.~O.~Barvinsky and A.~Y.~Kamenshchik,
Quantum scale of inflation and particle physics of the early universe,
Phys.~Lett.~B {\bf 332}, 270 (1994).

\bibitem{we10}
A.~O.~Barvinsky, A.~Y.~Kamenshchik and I.~V.~Mishakov,
Quantum origin of the early inflationary universe,
Nucl.~Phys.~{\bf B491}, 387 (1997).

\bibitem{we2}
A.~O.~Barvinsky and A.~Y.~Kamenshchik,
Effective equations of motion and initial conditions for inflation in
quantum cosmology,
Nucl.~Phys.~{\bf B532}, 339 (1998).


\bibitem{Shap}
F.~L.~Bezrukov and M.~Shaposhnikov,
The standard model Higgs boson as the inflaton,
Phys.~Lett.~B {\bf 659}, 703 (2008).

\bibitem{BKS}
A.~O.~Barvinsky, A.~Y.~Kamenshchik and A.~A.~Starobinsky,
Inflation scenario via the standard model Higgs boson and LHC,
J.~Cosmol.~Astropart.~Phys.~11 (2008) 021.

\bibitem{Shap1}
F.~L.~Bezrukov, A.~Magnin and M.~Shaposhnikov,
Standard model Higgs boson mass from inflation,
Phys~Lett.~B {\bf 675}, 88 (2009).

\bibitem{Shap2}
F.~Bezrukov and M.~Shaposhnikov,
Standard model Higgs boson mass from inflation: two loop analysis,
J.~High Energy Phys.~07 (2009) 089.

\bibitem{Wil}
A.~De Simone, M.~P.~Hertzberg and F.~Wilczek,
Running Inflation in the standard model,
Phys.~Lett.~B {\bf 678}, 1 (2009).

\bibitem{BKKSS}
A.~O.~Barvinsky, A.~Y.~Kamenshchik, C.~Kiefer, A.~A.~Starobinsky and C.~Steinwachs,
Asymptotic freedom in inflationary cosmology with a nonminimally coupled Higgs field,
J.~Cosmol.~Astropart.~Phys.~12 (2009) 003.

\bibitem{Clark}
T.~E.~Clark, B.~Liu, S.~T.~Love and T.~ter Veldhuis,
The standard model Higgs Boson-inflaton and dark matter,
Phys~Rev~D {\bf 80}, 075019 (2009).

\bibitem{Lerner}
R.~N.~Lerner and J.~McDonald,
Gauge singlet scalar as inflaton and thermal relic dark matter,
Phys.~Rev.~D {\bf 80}, 123507 (2009).

\bibitem{supergravity}
M.~B.~Einhorn and D.~R.~T.~Jones, Inflation with non-minimal gravitational couplings in supergravity,
J.~High Energy Phys.~03 (2010) 026.

\bibitem{supergravity1}
S.~Ferrara, R.~Kallosh, A.~Linde, A.~Marrani and A.~Van Proeyen,
Jordan Frame Supergravity and Inflation in NMSSM,
Phys.~Rev.~D {\bf 82}, 045003 (2010).

\bibitem{ShapNew}
F.~Bezrukov, M.~Yu.~Kalmykov, B.~A.~Kniehl and M.~Shaposhnikov,
Higgs boson mass and new physics,
J.~High Energy Phys.~10 (2012) 140.

\bibitem{LindeNew}
R.~Kallosh and A.~Linde,
Superconformal Generalization of the Chaotic Inflation Model $\lambda\,\phi^4/4-\xi/2\,\phi^2\,R$,
J.~Cosmol.~Astropart.~Phys.~06 (2013) 027.

 
\bibitem{AtlasResultHiggs}
G.~Aad et al. [ATLAS Collaboration],
Observation of a new particle in the search for the standard model Higgs boson with the ATLAS detector at the LHC,
Phys.~Lett.~B, {\bf 716}, 1 (2012).

\bibitem{CMSResultHiggs}
S.~Chatrchyan et al. [CMS Collaboration],
Observation of a new boson at a mass of 125 GeV with the CMS experiment at the LHC,
Phys.~Lett.~B, {\bf 716}, 30 (2012).

\bibitem{Planck}
P.~A.~R.~Ade et al. [Planck Collaboration],
Planck 2013 results. XXII. Constraints on inflation,
Astron.~Astrophys.~571, A22 (2014). 


\bibitem{Kaiser}
D.~I.~Kaiser,
Conformal transformations with multiple scalar fields,
Phys.~Rev.~D {\bf 81}, 084044 (2010).

\bibitem{LernerII}
R.~N.~Lerner, J.~McDonald,
A unitarity-conserving Higgs inflation model,
J.~Cosmol.~Astropart.~Phys.~04 (2010) 015.

\bibitem{Hertzberg}
M.~P.~Hertzberg,
On inflation with non-minimal coupling,
J.~High Energy Phys.~11 (2010) 023.


\bibitem{Burgess1}
C.~P.~Burgess, H.~Lee, M.~Hyun and M.~Trott,
Power-counting and the validity of the classical approximation during inflation
J.~High Energy Phys.~09 (2009) 103.

\bibitem{Espinoza}
J.~L.~F.~Barbon and J.~R.~Espinosa,
On the naturalness of Higgs inflation,
Phys.~Rev.~D {\bf 79}, 081302  (2009).

\bibitem{Burgess2}
C.~P.~Burgess, H.~M.~Lee, and M.~Trott,
Comment on Higgs inflation and naturalness,
J.~High Energy Phys.~07 (2010) 007.

\bibitem{Lerner1}
R.~N.~Lerner and J.~McDonald,
Higgs inflation and naturalness,
J.~Cosmol.~Astropart.~Phys.~04 (2010) 015.

\bibitem{Giudice}
G.~F.~Giudice and H.~M.~Lee,
Unitarizing Higgs inflation,
Phys.~Lett.~B {\bf 694}, 294 (2011).

\bibitem{BMSS}
F.~Bezrukov, A.~Magnin, M.~Shaposhnikov and S.~Sibiryakov,
Higgs inflation: consistency and generalisations,
J.~High Energy Phys.~01 (2011) 016.

\bibitem{BKKSS1}
A.~O.~Barvinsky, A.~Y.~Kamenshchik, C.~Kiefer, A.~A.~Starobinsky and C.~F.~Steinwachs,
Higgs boson, renormalization group, and cosmology,
Eur.~Phys.~J.~C {\bf 72}, 2219 (2012).

\bibitem{Lee}
H.~M.~Lee, Running inflation with unitary Higgs, 
Phys.~Lett.~B {\bf 722}, 198 (2013).

\bibitem{Calmet:2013hia}
 X.~Calmet and R.~Casadio,
 Self-healing of unitarity in Higgs inflation,
 Phys.~Lett.~B {\bf 734}, 17 (2014).
  

\bibitem{Faraoni}
V.~Faraoni and E.~Gunzig,
Einstein frame or Jordan frame?,
Int.~J.~Theor.~Phys.~{\bf 38}, 217 (1999).

\bibitem{Capozziello2}
S.~Capozziello, P.~Martin-Moruno, C.~Rubano,
Physical nonequivalence of the Jordan and Einstein frames,
Phys.~Lett.~B {\bf 689}, 117 (2010).

\bibitem{Capozziello1}
S.~Capozziello. and D.~Saez-Gomez,
Conformal frames and the validity of Birkhoffs theorem,
AIP Conf.~Proc. {\bf 1458}, 347 (2011).

\bibitem{Sasaki}
 N.~Deruelle and M.~Sasaki,
 Conformal equivalence in classical gravity: the example of ``veiled''general relativity,
 Springer Proc.~Phys.~{\bf 137}, 247 (2011). 
  
\bibitem{WeProc}
C.~F.~Steinwachs and A.~Yu.~Kamenshchik,
Non-minimal Higgs-inflation and frame dependence in vcosmology,
AIP Conf.~Proc.~{\bf 1514}, 161 (2012).

\bibitem{Calmet}
X.~Calmet and T.~Yang,
Frame transformations of gravitational theories,
Int.~J.~Mod.~Phys.~{\bf A28}, 1350042 (2013).

\bibitem{Postma:2014vaa}
 M.~Postma and M.~Volponi,
 Equivalence of the Einstein and Jordan frames,
Phys.~Rev.~D {\bf 90}, 102516 (2014).
  
\bibitem{Obukhov:2014mka}
  Y.~N.~Obukhov and D.~Puetzfeld,
 Equations of motion in scalar-tensor theories of gravity: A covariant multipolar approach,
  Phys.~Rev.~D {\bf 90}, 10, 104041 (2014).

\bibitem{Domenech:2015qoa}
  G.~Domènech and M.~Sasaki,
  Conformal frame dependence of inflation,
  arXiv:1501.07699.
    
  
\bibitem{Catena:2006bd}
  R.~Catena, M.~Pietroni and L.~Scarabello,
 Einstein and Jordan reconciled: a frame-invariant approach to scalar-tensor cosmology,
  Phys.~Rev.~D {\bf 76}, 084039 (2007). 
  
\bibitem{Kaiser:2010ps}
  D.~I.~Kaiser,
 Conformal transformations with multiple scalar fields,
  Phys.~Rev.~D {\bf 81}, 084044 (2010). 
  
  \bibitem{Kubota:2011re}
  T.~Kubota, N.~Misumi, W.~Naylor and N.~Okuda,
 The conformal transformation in general single field inflation with nonminimal coupling,
  J.~Cosmol.~Astropart.~Phys.~02 (2012) 034.
  
 \bibitem{Prokopec:2012ug}
  T.~Prokopec and J.~Weenink,
  Uniqueness of the gauge invariant action for cosmological perturbations,
  J.~Cosmol.~Astropart.~Phys.~12 (2012) 031.
  
 \bibitem{Prokopec:2013zya}
  T.~Prokopec and J.~Weenink,
  Frame independent cosmological perturbations,
  J.~Cosmol.~Astropart.~Phys.~09 (2013) 027. 
  
\bibitem{Chiba:2013mha}
  T.~Chiba and M.~Yamaguchi,
  Conformal-frame (in)dependence of cosmological observations in scalar-tensor theory,
  J.~Cosmol.~Astropart.~Phys.~10 (2013) 040. 
  
\bibitem{Prokopec:2014iya}
  T.~Prokopec and J.~Weenink,
  Naturalness in Higgs inflation in a frame independent formalism,
  arXiv:1403.3219.
  

\bibitem{we-paper1}
 C.~F.~Steinwachs and A.~Yu.~Kamenshchik,
One-loop divergences for the gravity nonminimally coupled with a scalar field multiplet:
calculation in the Jordan frame. I. The main results,
Phys.~Rev.~D {\bf 84}, 024026 (2011).

\bibitem{DeWitt}
B.~S.~DeWitt,
{\it Dynamical theory of groups and fields} (Gordon and Breach, New York, 1965).

\bibitem{Bar-Vil}
A.~O.~Barvinsky and G.~A.~Vilkovisky,
The Generalized Schwinger-Dewitt Technique In Gauge Theories And Quantum
Gravity,
Phys.~Rep.~{\bf 119}, 1 (1985).

  

\bibitem{Borchers}
H.~I.~Borchers,
\"Uber die Mannigfaltigkeit der interpolierenden Felder zu einer kausalen S-Matrix,
Nuovo Cim. {\bf 15}, 784 (1960).

\bibitem{Chisholm}
S.~R.~Chisholm,
Change of variables in quantum field theories,
Nucl.~Phys.~{\bf 26}, 469 (1961).

\bibitem{Kamefuchi}
  S.~Kamefuchi, L.~O'Raifeartaigh and A.~Salam,
  Change of variables and equivalence theorems in quantum field theories,
  Nucl.~ Phys.~{\bf 28}, 529 (1961).

\bibitem{Coleman}
 S.~Coleman, J.~Wess and B.~Zumino,
 Structure of phenomenological lagrangians. I,
 Phys.~Rev.~{\bf 177}, 2239 (1969).

\bibitem{Kallosh}
R.~E.~Kallosh and I.~V.~Tyutin,
Yad.~Fiz.~{\bf 17}, 190 (1973) [The equivalence theorem and gauge invariance in renormalizable theories, Sov.~J.~Nucl.~Phys.~{\bf 17}, 98 (1973)].

\bibitem{Tyutin}
I.~V.~Tyutin, Yad.~Fiz.~{\bf 35}, 222-228 (1982)[
Parametrization dependence of nonlinear quantum field theories,
Sov.~J.~Nucl.~Phys.~{\bf35}, 125 (1982)].

\bibitem{Tyutin:2000ht}
  I.~V.~Tyutin, Yad.~Fiz.~{\bf 65}, 201 (2002)[
 Once again on the equivalence theorem,
  Phys.~At.~Nucl.~{\bf 65}, 194 (2002)].
  

\bibitem{MathTensor}
L.~Parker and S.~M.~Christensen, 
{\it MathTensor: A system for doing tensor analysis by computer},
(Addison-Wesley, Redwood City, 1994).

\bibitem{Shapiro}
I.~L. Shapiro and H.~Takata,
One loop renormalization of the four-dimensional theory for quantum dilaton gravity,
Phys.~Rev.~D {\bf 52}, 2162 (1995).

\bibitem{Alvarez:2014qca}
  E.~Alvarez, M.~Herrero-Valea and C.~P.~Martin, Conformal and nonconformal dilaton gravity,
  J.~High Energy Phys.~10 (2014) 115.
  
\bibitem{Bamba:2014mua} 
  K.~Bamba, G.~Cognola, S.~D.~Odintsov and S.~Zerbini,
 One-loop modified gravity in de Sitter universe, quantum corrected inflation, and its confrontation with the Planck result,
  Phys.~Rev.~D {\bf 90}, 023525 (2014).
 
\bibitem{Nojiri:2000ja}
 S.~Nojiri and S.~D.~Odintsov,
 Quantum dilatonic gravity in (D = 2) dimensions, (D = 4) dimensions and (D = 5) dimensions,
  Int.~J.~Mod.~Phys.~{\bf A16}, 1015 (2001).  
  
\bibitem{we-renorm}
A.~O.~Barvinsky, A.~Y.~Kamenshchik and I.~P.~Karmazin,
The renormalization group for nonrenormalizable theories: Einstein gravity
with a scalar field,
Phys.~Rev.~D {\bf 48}, 3677 (1993).
  


\bibitem{George:2013iia}
  D.~P.~George, S.~Mooij and M.~Postma,
  Quantum corrections in Higgs inflation: the real scalar case,
  J.~Cosmol.~Astropart.~Phys.~02 (2014) 024.

\bibitem{Vilkovisky}
  G.~A.~Vilkovisky,
  The unique effective action in quantum field theory,
  Nucl.~Phys.~B {\bf 234}, 125 (1984).  
  
\bibitem{WeNew}
A.~Yu.~Kamenshchik and C.~F.~Steinwachs, work in progress.

\bibitem{Kamenshchik:2013dga}
  A.~Y.~Kamenshchik, E.~O.~Pozdeeva, A.~Tronconi, G.~Venturi and S.~Y.~Vernov,
  Integrable cosmological models with nonminimally coupled scalar fields,
  Classical Quantum Gravity~{\bf 31}, 105003 (2014). 
  
\bibitem{Fre}
  P.~Fr\'e, A.~Sagnotti and A.~S.~Sorin,
   Integrable scalar cosmologies I. Foundations and links with string theory,
   Nucl.~Phys.~{\bf B877}, 1028 (2013).  

\bibitem{Moss:2014nya} 
  I.~G.~Moss,
 Covariant one-loop quantum gravity and Higgs inflation,
  arXiv:1409.2108.

\bibitem{WeTunneling}
A.~O.~Barvinsky, A.~Yu.~Kamenshchik, C.~Kiefer and C.~F.~Steinwachs,
Tunneling cosmological state revisited: origin of inflation with a nonminimally coupled standard model Higgs inflaton,
Phys.~Rev.~D {\bf 81}, 043530 (2010).

\bibitem{Calcagni:2014xca}
  G.~Calcagni, C.~Kiefer and C.~F.~Steinwachs,
 Quantum cosmological consistency condition for inflation,
 J.~Cosmol.Astropart.~Phys.~10 (2014) 026.

\bibitem{Fradkin}
E.~S.~Fradkin and G.~A.~Vilkovisky,
On renormalization of quantum field theory in curved space-time,
Lett.~Nuovo Cimento Soc. Ital. Fis.~{\bf 19}, 47 (1977).
\end{thebibliography}
\end{document}